\documentclass[%
 reprint,
 amsmath,amssymb,
 aps,
]{revtex4-2}

\usepackage[colorlinks = true,
            linkcolor = blue,
            urlcolor  = blue,
            citecolor = blue,
            anchorcolor = blue]{hyperref}
\usepackage{nameref}
\usepackage{graphicx}
\usepackage{dcolumn}
\usepackage{bm}
\usepackage{orcidlink}
\newcommand{\aap}{A\&A}

\newcommand{\mnras}{MNRAS}

\newcommand{\aapr}{AApR}

\newcommand{\jgr}{Journal of Geophysical Research}
\newcommand{\apjl}{The Astrophysical Journal Letters}

\usepackage{ulem}
\usepackage{xcolor}

\newcommand{\TRISTAN}{{\bf \textsf{\small Tristan-MP}} }

\newcommand{\alfven}{Alfv\'{e}n }

\newcommand{\Ma}{\mathcal{M}_{\rm A}}
\newcommand{\Ms}{\mathcal{M}_{\rm s}}
\newcommand{\Mse}{\mathcal{M}_{\rm s,e}}
\newcommand{\mr}{m_{\rm R}}
\newcommand{\thetabn}{\theta_{\rm Bn}}

\begin{document}

\preprint{APS/123-QED}

\title{Speed-dependent Threshold for Electron Injection into Diffusive Shock Acceleration}
\author{Siddhartha Gupta$^1$\,\orcidlink{0000-0002-1030-8012}}
\email{gsiddhartha@princeton.edu}
\author{Damiano Caprioli$^{2,3}$\,\orcidlink{0000-0003-0939-8775}}
\author{Anatoly Spitkovsky$^1$\,\orcidlink{0000-0001-9179-9054}}
\affiliation{$^1$Department of Astrophysical Sciences, Princeton University, 4 Ivy Lane, Princeton, NJ 08544, USA}
\affiliation{$^2$Department of Astronomy and Astrophysics, The University of Chicago, IL 60637, USA}
\affiliation{$^3$Enrico Fermi Institute, The University of Chicago, Chicago, IL 60637, USA}
\date{\today}

\begin{abstract}
Finding the injection threshold for diffusive shock acceleration (DSA) of electrons in collisionless shocks has been a longstanding unsolved problem. 
Using first-principles kinetic simulations, we identify the conditions for electron injection into DSA and quantify the evolution of the nonthermal tail in self-generated electromagnetic turbulence. 
By analyzing electron trajectories and their momentum gain during shock-recrossing cycles, we demonstrate that electrons start participating in DSA when their speed is large enough to overrun the shock. 
We develop a minimal model showing that speed-dependent injection reproduces nonthermal electron spectra observed in kinetic simulations. 
Our findings establish a new criterion for electron DSA, which has broad implications for the nonthermal emission of shock-powered space/astrophysical systems.
\end{abstract}

\maketitle

\section{Introduction}\label{sec:intro}
%
Nonthermal (NT) electrons power a significant fraction of the observed radiation in radio, X-ray, and $\gamma$-ray wavelengths from space/astrophysical sources. 
While shocks are prime sites for producing these NT electrons, the precise mechanisms that inject them from a thermal background into the NT state have remained elusive, despite decades of research \citep[see][for reviews]{caprioli15p, marcowith+16, bohdan22}.
Uncovering the electron injection criteria is fundamental to understanding NT energy production.

The widely-accepted diffusive shock acceleration (DSA) mechanism suggests that particles repeatedly crossing a strong shock gain energy, yielding the power-law momentum distribution, $f(p)\propto p^{-4}$ \citep[][]{krymskii77,axford+77p,blandford+78,bell78a} needed to explain NT emission in many Galactic and extra-galactic sources \citep[][]{vanweeren+10,su+10,morlino+12,SN1006HESS,masters+16,liu+19,jebaraj+24,giuffrida+22}.
DSA is considered a universal mechanism because the slope results from the balance between two key ingredients:
1) the momentum gain per acceleration cycle and
2) the probability of escape from the accelerating region, both of which depend on the shock compression ratio \cite{bell78a}.
However, \textit{how many} particles enter the DSA process is a much more complicated problem. 

The general intuition is that particles must have Larmor radii exceeding the shock thickness to be injected into DSA \citep[][]{blasi+05, malkov+01, kang+02}.
This condition is easily met by protons and heavier ions participating in DSA because the shock thickness is determined by the proton dynamics and is comparable to the Larmor radius of thermal protons, $R_{\rm L,i} \sim (p/m_{\rm i}v_{\rm A}) d_{\rm i}$,  where $p$ is the particle momentum, $m_{\rm i}$ is the proton mass, $d_{\rm i}$ is the proton skin depth and $v_{\rm A}$ is the \alfven speed \citep[][]{treumann09, caprioli+15}.
However, first-principles kinetic simulations \citep[e.g.,][]{caprioli+14a, park+15, kumar+21} show that the Larmor radius condition does not directly govern injection.
Instead, proton injection depends on whether they can outrun the shock after being reflected, which for quasi-parallel shocks corresponds to a speed threshold of $v_{\rm inj} \gtrsim 3v_{\rm sh}$, where $v_{\rm sh}$ is the shock velocity \citep[][]{caprioli+15}.
At this speed, protons have a momentum $p_{\rm inj}$ and Larmor radius $R_{\rm L,i} \approx 3 \Ma d_{\rm i}$, where $\Ma=v_{\rm sh}/v_{\rm A}$ is the \alfven Mach number, that exceeds the shock thickness, making the speed condition indistinguishable from the Larmor radius requirement. 

Although speed- and momentum-dependent criteria are equivalent for ions, they are not the same for electrons \citep[][]{bohdan22,amano+22}.
Electrons with momentum above $p_{\rm inj}\sim 3m_{\rm i} v_{\rm sh}$ clearly participate in DSA \citep[][]{park+15}; however, can lower-momentum electrons be involved in DSA, too?
Electrons would need significant pre-acceleration to meet the proton momentum threshold, which poses major challenges \citep[e.g.,][]{ball+01,hoshino+02,amano+07,riquelme+11,matsumoto+12,guo+14a,kato15,park+15,crumley+19,bohdan+19a,xu+20,arbutina+21,kumar+21,shalaby+22,morris+22,zekovic+24}, while a speed-dependent threshold demands much less energization.

In this Letter, we address the electron DSA injection problem by investigating electron trajectories in fully kinetic Particle-In-Cell (PIC) simulations. 
We examine when electrons outrun the shock, track their instances of energy gain and compare these with predictions from DSA theory.
We provide -- for the first time -- conclusive evidence of speed-dependent injection and develop a simulation-validated minimal model to illustrate the necessary and sufficient steps of electron DSA.

The main difference between electron and ion injection lies in their dynamics in the shock foot.
Since reflected ions have Larmor radii comparable to the shock transition width, they are generally not tied to any particular field line and experience scattering and energy gains on every gyration via shock-drift acceleration.
This process is ``lossy", in the sense that at any shock encounter a fraction of $\mathcal{O}(1)$ of the particles is lost downstream.
This proceeds until the remaining ions reach the injection velocity and can outrun the shock along the upstream field lines, entering DSA \citep[][]{caprioli15}.

In contrast, electrons have much smaller Larmor radii, generally follow the field lines in the shock foot, and are unlikely to experience scattering within a single gyration. 
Electrons that arrive at the shock sufficiently preheated can reflect (or ``mirror'') off the shock ramp. 
Such reflected electrons then stream along the magnetic field and scatter (isotropize) due to turbulence generated by shock-returning particles. 
This process allows them to re-approach the shock with different pitch angles and mirror again -- a lossy process that leads to both fractional energy gain and escape probability similar to DSA\citep[][]{bell78a}.
The threshold for electron injection is then governed not by their Larmor radius, but by having sufficient speed to mirror off the shock and outrun it along the magnetic field. Thus, as we show in this paper, electron DSA spectra start from much smaller momenta than $p_{\rm inj}$.

The observational evidence that in both shock-powered sources \cite{morlino+12, brunetti+14, diesing+23} and Galactic cosmic rays \cite{pamela11, ams19b} the electron/proton ratio $K_{\rm ep}\ll 1$ at relativistic energies is recovered because: 1) the electron tail starts at lower momenta than the ions' and 2) to achieve the same momentum electrons need to undergo a larger number of lossy shock encounters.
\section{Kinetic Simulations and Results}\label{sec:results}
Using a fully kinetic PIC code \TRISTAN \citep[][]{buneman93,spitkovsky05}, we conduct nonrelativistic shock simulation in the upstream rest frame filled with equal-temperature ($T_{\rm e} = T_{\rm i}$) electron-proton thermal plasma \citep[as detailed in][]{gupta+24b}. 
We launch the shock using a reflecting wall (piston) moving at a speed $v_{\rm pt} = 0.1c$ along the $x$-direction.
The initial magnetic field is characterized by the \alfven speed $v_{\rm A}$ and is inclined relative to the shock normal at an angle $\thetabn = 30^\circ$, which is later reoriented by proton-driven turbulence \citep[][]{caprioli+14b,park+15,gupta+24b}. 
We parameterize our simulations by the shock speed $v_{\rm sh}=v_{\rm pt}\mathcal{R}/(\mathcal{R}-1)$, with the density compression ratio  $\mathcal{R}\approx 4$, \alfven Mach number $\Ma =v_{\rm sh}/v_{\rm A}$, proton sonic Mach number $\Ms  =v_{\rm sh}/v_{\rm th,i}$ (where $v_{\rm th,i}=\sqrt{k_{\rm B}T_{\rm i}/m_{\rm i}}$ is the proton thermal speed), and proton-to-electron mass ratio $\mr =m_{\rm i}/m_{\rm e}$. 
The corresponding sonic Mach number for electrons is $\Mse =\Ms  /\sqrt{\mr }$. We analyze high ($\Ma  = 20$) and low ($\Ma  = 5$) Mach shocks at fixed $\Ms=40$ and $\mr=100$, running until the system achieves a quasi-stationary state, at $\sim 300\omega_{\rm ci}^{-1}$, where $ \omega_{\rm ci}$ is the upstream ion gyrofrequency. 
We have validated our findings against different values of $\mr$, space resolution, and number of particles per cell.
\subsection{High Alfv\'{e}n Mach Number Shock}\label{subsec:highMAshock}
%
\begin{figure}[ht!]
    \centering
    \includegraphics[width=3.25in]{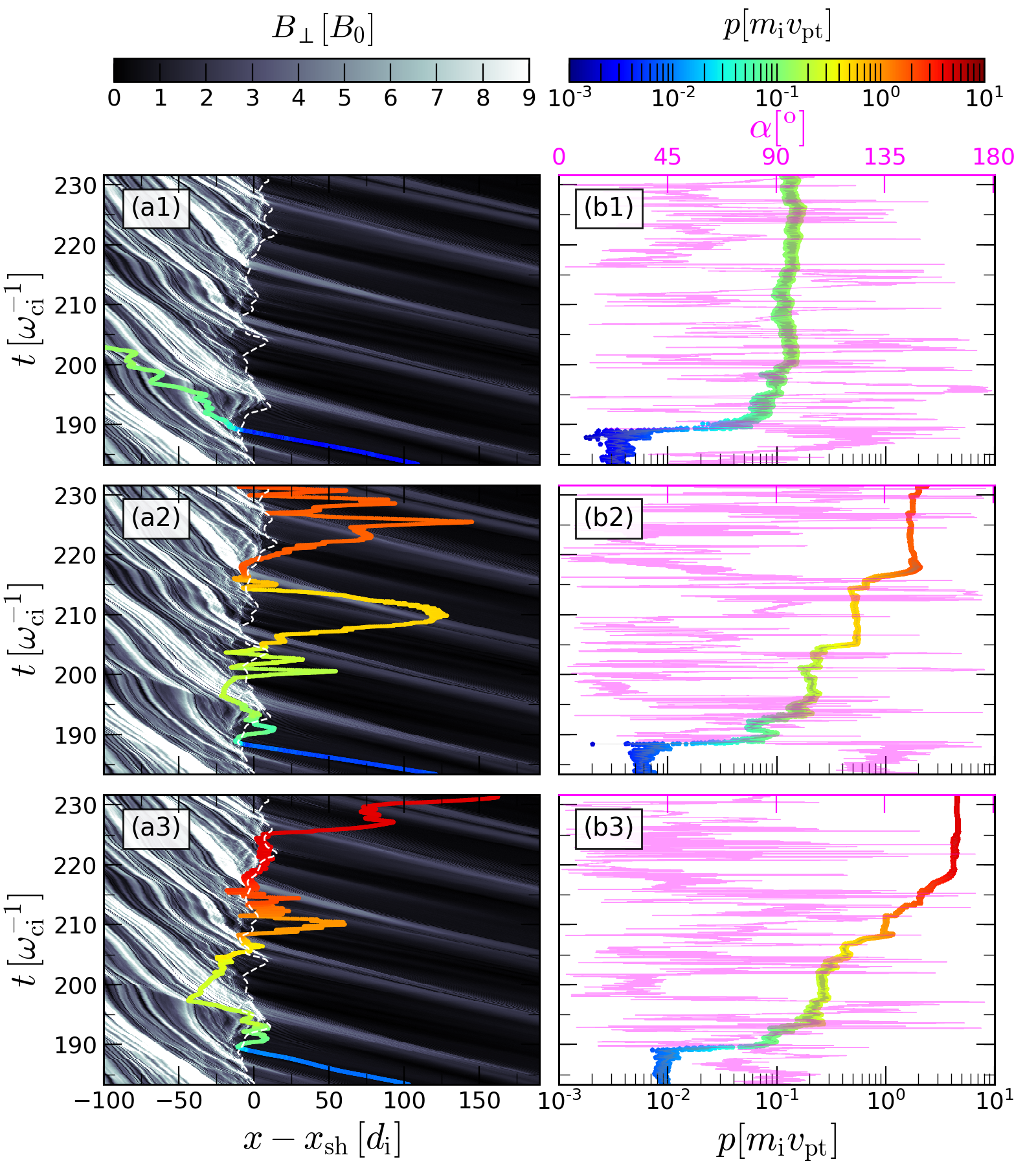}
    \caption{
    Typical electron trajectories in a shock with $\Ma =20$, $\Ms  =40$, and $\mr =100$.
    The background colors display the evolution of the transverse field $B_{\rm \perp}=\sqrt{B_{\rm y}^2+B_{\rm z}^2}$; the horizontal axis denotes the mean distance from the shock, with \textit{instantaneous} shock locations denoted by the dashed white curves.
    Trajectories are color-coded with momentum $p$ (in the upstream frame), also shown on the bottom axis of panels b;  
    the local pitch angle $\alpha$ between momentum and magnetic field is shown by magenta curves (top axis).
    }
    \label{fig:M20trajectory}
\end{figure}
For the general structure of the shock, we refer to the extensive description of our benchmark run in \citep{gupta+24b};
here we focus on selected particle trajectories. 

Fig.~\ref{fig:M20trajectory} shows the trajectory of three representative electrons in a $\Ma =20$ shock.
Figs.~\ref{fig:M20trajectory}(a1) and  \ref{fig:M20trajectory}(b1) display the typical trajectory of an electron that ends up in the downstream thermal population.
While upstream, its momentum remains almost constant at $p\sim m_{\rm e}v_{\rm pt}<10^{-2}m_{\rm i}v_{\rm pt}$, though its pitch angle, $\alpha$, fluctuates rapidly (magenta curve in Fig.~\ref{fig:M20trajectory}(b1)). 
When crossing the shock, its momentum increases due to fluctuating fields in the shock transition; 
such a heating favors electron/ion thermal equilibration \citep[][]{gupta+24b}.

Figs.~\ref{fig:M20trajectory}(a2) and (b2) and Figs.~\ref{fig:M20trajectory}(a3) and (b3) show electrons that are efficiently accelerated. 
They start with momenta similar to the electron in Figs.~\ref{fig:M20trajectory} (a1) and become suprathermal ($p/m_{\rm i}v_{\rm pt}\sim 10^{-2}$) as they approach the shock, consistent with the heating expected due to the instability generated by specularly-reflected protons \citep[][]{hoshino+02, muschietti+17,gupta+24a, lichko+24}. 
For the simulation parameters ($m_{\rm R}=100$, $v_{\rm pt}/c=0.1$), the corresponding electron speed is $u \sim 0.1 c$, comparable to the shock speed in the upstream frame, $v_{\rm sh}=0.133\,c$.
Unlike the previous one, these electrons are reflected by the shock, propagate upstream more than one gyroradius, and repeatedly come back to the shock, a clear signature of DSA.

The pitch angle evolution of NT electrons (magenta curves in Figs.~\ref{fig:M20trajectory}(b2)--\ref{fig:M20trajectory}(b3)) suggests that electrons are effectively scattered throughout their acceleration process, which both favors isotropization in the local wave frame and diffusion back to the shock, two clear signatures of DSA.
\subsection{DSA Injection Threshold}\label{subsec:injecth}
%
To further support the idea that electrons can undergo DSA as soon as they are able to outrun the shock, i.e., that they are injected when they reach a threshold in velocity, we measure the fractional momentum change $\Delta p/p$ in one complete (upstream--downstream--upstream) cycle and compare it with the DSA prediction, a procedure similar to the one used for protons \citep[][]{caprioli+20}.

\begin{figure}[t!]
    \centering
    \includegraphics[width=0.45\textwidth, trim=2px 0px 2px 0px, clip=true]{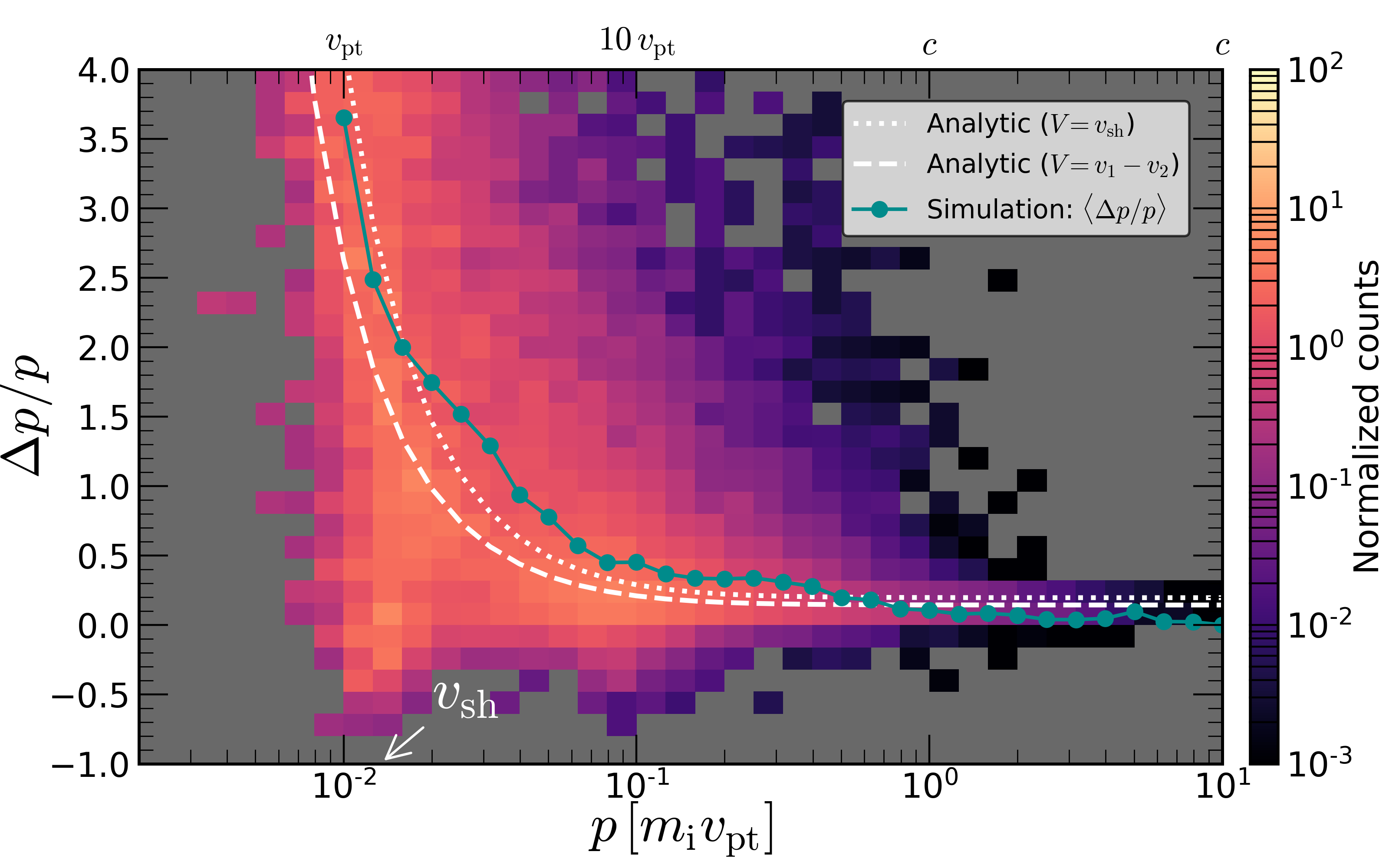}
    \caption{
    Fractional momentum change $\Delta p/p$ in one upstream--downstream--upstream cycle, as a function of speed/momentum (top/bottom axis) for the $\Ma =20$ shock.
    The cyan and white curves show the mean of the distribution obtained from our simulations and Eq. (\ref{eq:frac_gainp}), respectively.
    The measured $\langle \Delta p/p\rangle$ matches the DSA prediction when electron speed exceeds the shock speed, shown by the arrow.
    }
    \label{fig:M20gain}
\end{figure}

Fig.~\ref{fig:M20gain} shows the distribution of $\Delta p/p$ as a function of $p$ obtained from the analysis of about $2\times 10^{5}$ electron tracks, as detailed in Appendix \ref{app:A}.
The spread in $\Delta p/p$ arises because it depends on the particle's ingoing/outgoing directions.  
Averaging over all the particle trajectories returns the cyan line, $\langle\Delta p/p\rangle(p)$, compared with the analytic expectations for both non-relativistic and relativistic particles  (white curves), which read:
\begin{equation}\label{eq:frac_gainp}
    \mathcal{G} \equiv \left\langle\frac{\Delta p}{p}\right\rangle\approx \frac{2}{3}\frac{V}{u}\left[1+\sqrt{1+\left(\frac{V}{u}\right)^2}+\frac{3}{2}\frac{V}{u}\right]. 
\end{equation}
Here, $u=p/\gamma$ ($\gamma$ is the Lorentz factor) is the particle speed and $V$ is the relative speed between the scattering centers that confine particles at the shock. 
This can be either the relative speed between upstream and downstream $v_{\rm pt}=v_{\rm 1}-v_{\rm 2}$ (dashed white curve) or  $V\equiv v_{\rm 1}=v_{\rm sh}$ (dotted white curve), depending on whether electrons are scattered off the shock or off the downstream medium.
In the limit $u\rightarrow c$, choosing $V=|v_{\rm 1}-v_{\rm 2}|\equiv v_{\rm pt}$ returns $\mathcal{G}=4 V/3c$, the classical DSA prediction \citep[e.g.,][]{bell78a}. 
In the $V=v_{\rm sh}$ scenario, isotropization occurs at shock front and $\mathcal{G}$ is slightly larger.

Fig.~\ref{fig:M20gain} suggests that all electrons with $u>v_{\rm sh}$ gain momentum as predicted by Eq. (\ref{eq:frac_gainp}).
The slight preference of the measured $\langle \Delta p / p \rangle$ for the curve $V = v_{\rm sh}$ 
indicates that electrons return from the shock. 
Note that the prediction assumes that particles are effectively isotropic in the fluid frame, so we also confirm that this hypothesis is realized for electrons.
Thus, we demonstrate that above a \textit{speed} (and not momentum) threshold electrons are injected into DSA and gain energy according to the expected $\langle \Delta p/p \rangle$.
\subsection{Low Alfv\'{e}n Mach Number Shock}\label{subsec:lowMAshock}
\begin{figure}[t]
    \centering
    \includegraphics[width=3.25in]{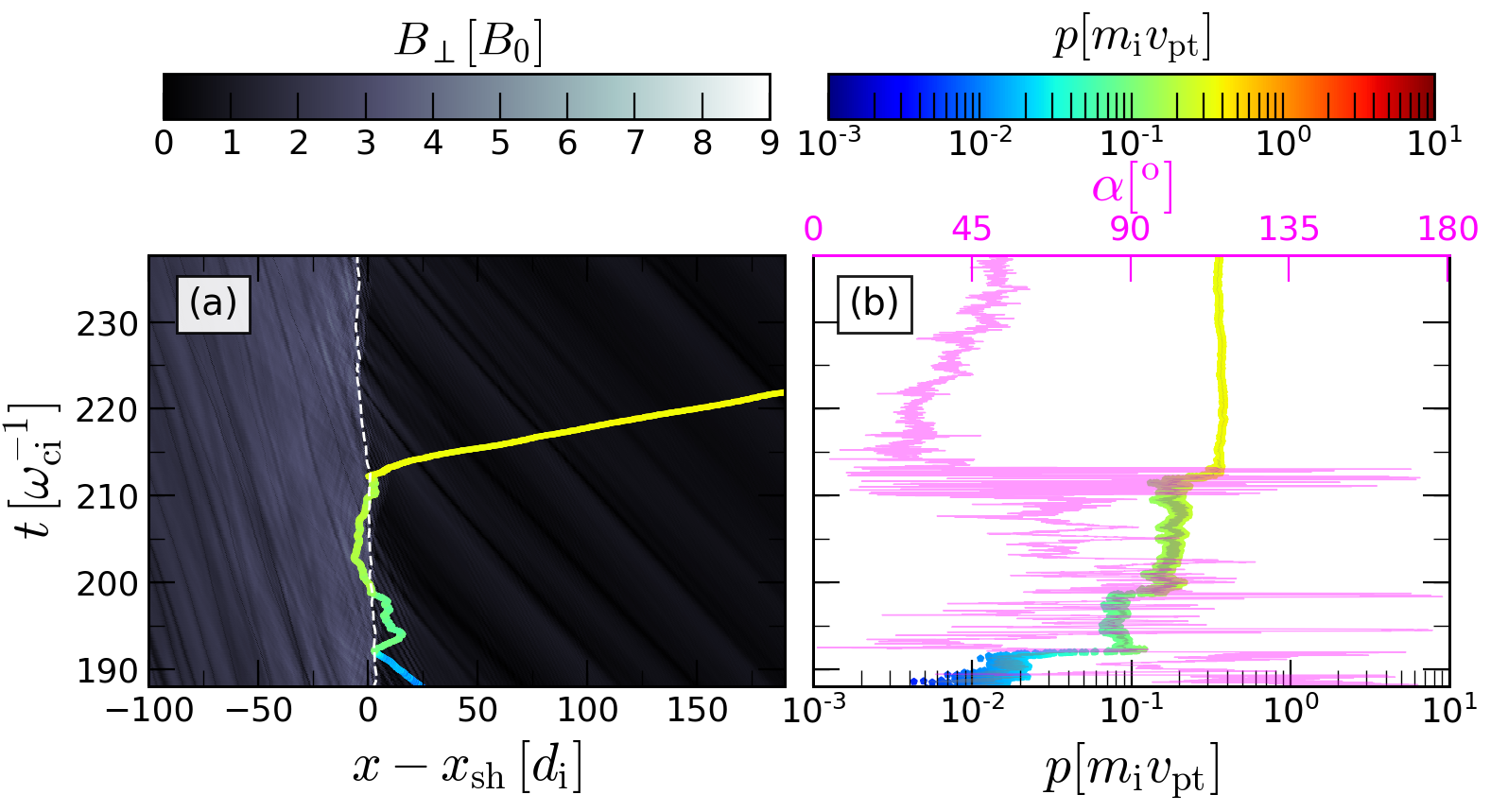}
    \caption{
     As in Fig.~\ref{fig:M20trajectory}, but for a shock with $\Ma =5$.}
    \label{fig:M5trajectory}
\end{figure} 

To test if this conclusion applies also to a low-$\Ma $ shock, we examine NT electron trajectories for a quasi-parallel $\Ma =5$ shock.
A representative electron in shown in Fig.~\ref{fig:M5trajectory}.
The background color in Fig.~\ref{fig:M5trajectory}(a) suggests that the upstream is less turbulent than in the $\Ma =20$ shock in Fig~\ref{fig:M20trajectory}. 
Fig.~\ref{fig:M5trajectory} (b) shows that beyond a certain momentum, the pitch-angle fluctuations of the back-streaming electrons are very small, which corresponds to a slow spatial diffusion and thus slow acceleration. 
Importantly, also in this case electrons outrun the shock when $u>v_{\rm sh}$, despite their Larmor radii being much smaller than in the $\Ma =20$ shock.
This confirms that DSA injection is governed by particle speed, not momentum, regardless of the shock Mach number.
\subsection{Confinement and Acceleration Rate}\label{subsec:maxp}
%
We now analyze the DSA rate by estimating the self-generated diffusion coefficient, and quantifying the maximum energy achievable at a given time in both high- and low-Mach shocks.
It is useful to describe the transport of particles with mass $m_{\rm \alpha}$ as a function of the Bohm diffusion coefficient (mean free path equal to the gyroradius $R_{\rm L,\alpha}$):
\begin{eqnarray}\label{eq:Bohmdiff}
    D_{\rm B, \alpha} & = & u_{\rm \alpha}\, R_{\rm L,\alpha}  \equiv 
 \frac{\tilde{p}^2}{\gamma_{\rm \alpha}}\frac{m_{\rm i}}{m_{\rm \alpha}}\frac{\Ma ^2}{\tilde{v}_{\rm sh}^{2}}\left(\frac{B}{B_{\rm 0}}\right)^{-1}\,v_{\rm A} d_{\rm i} \ ,
\end{eqnarray}
where $\gamma_{\rm \alpha} = [1+\tilde{p}^2\,(m_{\rm i}/m_{\rm \alpha})^2 \beta_{\rm pt}^2]^{1/2}$, $\tilde{p}=p_{\rm \alpha}/m_{\rm i}v_{\rm pt}$, and $\tilde{v}_{\rm sh}=(v_{\rm sh}/v_{\rm pt})=\mathcal{R}/(\mathcal{R}-1)\simeq 4/3$.
The maximum momentum for species $\alpha$ at a given time can be obtained by equating the acceleration time, $\tau_{\rm acc}(p)=[3/(v_{\rm 1}-v_{\rm 2})](D_{\rm \alpha,1}/v_{\rm 1}+D_{\rm \alpha, 2}/v_{\rm 2})$ \citep[][]{drury83, blasi+07} with the shock dynamical time, which returns:
\begin{eqnarray}\label{eq:pmax}
    \frac{p_{\rm max,\alpha}}{m_{\rm i}v_{\rm pt}} &\approx& \,\frac{\tilde{t}\beta_{\rm pt}/\tilde{D}_{\rm 1,\alpha}}{\frac{9\sqrt{2}}{4}(1+\frac{4}{D_{\rm 12}})} \sqrt{1+\sqrt{1+\left[\frac{9 \frac{m_{\rm \alpha}}{m_{\rm i}}(1+\frac{4}{D_{\rm 12}}) }{2\tilde{t}\beta_{\rm pt}^2/\tilde{D}_{\rm 1,\alpha} }\right]^2}}\, . 
\end{eqnarray}
Here $\tilde{t}=t \omega_{\rm ci}$, $\tilde{D}_{\rm 1,\alpha}=D_{\rm 1,\alpha}/D_{\rm B0,\alpha}$, and $D_{\rm 12}=D_{\rm \alpha,1}/D_{\rm \alpha,2}$ is the ratio of the upstream to the downstream diffusion coefficients. 
Eq. (\ref{eq:pmax}) suggests that the electron maximum momentum is smaller than  protons' unless $t \gg (\beta_{\rm pt}^{-2})\,\omega_{\rm ci}^{-1}$;
both species converge to the same $p_{\rm max}$ only after many ion cyclotron times \citep[][]{park+15,crumley+19,arbutina+21,gupta+24b}.

We measure the local effective $D_{\rm \alpha}(p)$ in our simulations as described in \citep[][]{caprioli+14c} and in Appendix \ref{app:diffusion}.
For $\Ma =20$ shock, substituting $D_{\rm 1,e}\approx D_{\rm B0,e}$ and $D_{\rm 2,e}=0.1 D_{\rm B0,e}$ into Eq. (\ref{eq:pmax}) gives $p_{\rm max,e}/m_{\rm i}v_{\rm pt}\approx 9$, in good agreement with the maximum momentum obtained in kinetic simulations (see also Fig.~\ref{fig:M20trajectory}(b)).
For $\Ma =5$, a similar estimate returns $p_{\rm max,e}/m_{\rm i}v_{\rm pt}\approx 0.9$, one order of magnitude smaller than for $\Ma =20$. 
This explains why NT electrons are poorly confined in Fig.~\ref{fig:M5trajectory}. 
 In low-Mach number shocks, where there is little field amplification upstream, acceleration beyond $p\sim m_{\rm i}v_{\rm pt}$ is generally slow and the development of power-law tails is hard to achieve for typical PIC run times. 
\section{A Minimal Model for Electron DSA}\label{subsec:mimalmodel}
%
We now construct a minimal model that sums up the essential processes of electron DSA ---heating, reflection, injection, and acceleration--- and returns the electron spectra from PIC simulations.

The first and necessary stages are heating and reflection, which prevents particles from transmitting through the shock \citep[][]{schwartz+83, krauss-varban+89,ball+01,mann+06,guo+14a, caprioli+15}.
The magnetic mirroring criteria (i.e., that in the frame where the electric field vanishes the parallel and perpendicular velocities $u^{\prime}_{\rm \parallel}<0$ and $u^{\prime}_{\rm \perp} >u^{\prime}_{\rm \parallel}/\sqrt{\mathcal{R}_{\rm B}-1}$, where $\mathcal{R}_{\rm B}$ is the magnetic field compression ratio in the shock ramp) require electrons to impinge on the shock with a speed exceeding
\begin{equation}\label{eq:condi-3}
    u_{\rm min} \equiv \frac{v_{\rm sh}\sec\thetabn}{\sqrt{\{(1-(v_{\rm sh}\sec\thetabn/c)^2\}(\mathcal{R}_{\rm B}-1) + 1}}\, 
\end{equation}
to be reflected (see Appendix \ref{app:syntheticshock}). 
To ensure that a non-negligible fraction of upstream electrons meets such condition, the thermal speed must satisfy $\sqrt{3} v_{\rm th,e} \gtrsim u_{\rm min}$. 
For quasi-parallel shocks ($\thetabn\lesssim 45^{\rm o}$), this condition demands an effective $\Mse  \lesssim 3$ immediately upstream of the shock.
If $\Mse \gg 3$ far upstream, which is the case for all strong shocks with $\Ms\gg 3 (m_{\rm i}/m_{\rm e})^{1/2}$, 
electron injection thus requires some non-adiabatic heating, either in the precursor or in the shock foot \citep[e.g.,][]{hoshino+02,bohdan+22,gupta+24a}.
Such heating is typically observed for strong quasi-parallel shocks, as discussed extensively in \citep[][]{gupta+24a}.
Our analysis shows that $\sim 10\%$ of the electrons approaching the shock can be reflected when $\Mse \lesssim 3$ (see Appendix \ref{app:syntheticshock}).
These reflected electrons rapidly isotropize in the turbulence at the shock foot, so only a fraction of them have speeds along the shock normal exceeding the shock speed and are injected into DSA ($\eta_{\rm inj}$).

We next use a Monte Carlo approach to generate synthetic electron spectra.
We consider two cases: one strong shock with $\Ma =20$ and  $\thetabn\approx 45^{\rm o}$, which is the average inclination in the presence of strong self-generated turbulence, and one weak shock with $\Ma =5$ and $\thetabn=30^{\rm o}$, as weak turbulence is less effective in altering the initial inclination.
We consider electrons with a thermal distribution with $\Mse =2.5$.
We then classify electrons as transmitted or reflected based on the conditions described above: those failing the reflection criteria are advected downstream and thermalized to the post-shock temperature predicted by the Rankine-Hugoniot jump conditions, while those meeting it return upstream and are isotropized (as observed in Figs.~\ref{fig:M20trajectory} and \ref{fig:M5trajectory}).
Electrons whose speed exceeds $v_{\rm sh}$ undergo DSA, where the probability of remaining in the acceleration region is taken as $\mathcal{P}_{\rm rem} \simeq 1 - \frac{4}{\mathcal{R}-1}\left(v_{\rm pt}/u\right)$ (Appendix \ref{appsec:escapefrac}). 
To account for the finite runtime of the PIC simulations, we apply an upper bound on the maximum energy of NT electrons (see \S \ref{subsec:maxp}).
\begin{figure}[ht!]
    \centering
    \includegraphics[width=0.45\textwidth, trim=5px 0px 24px 0px, clip=true]{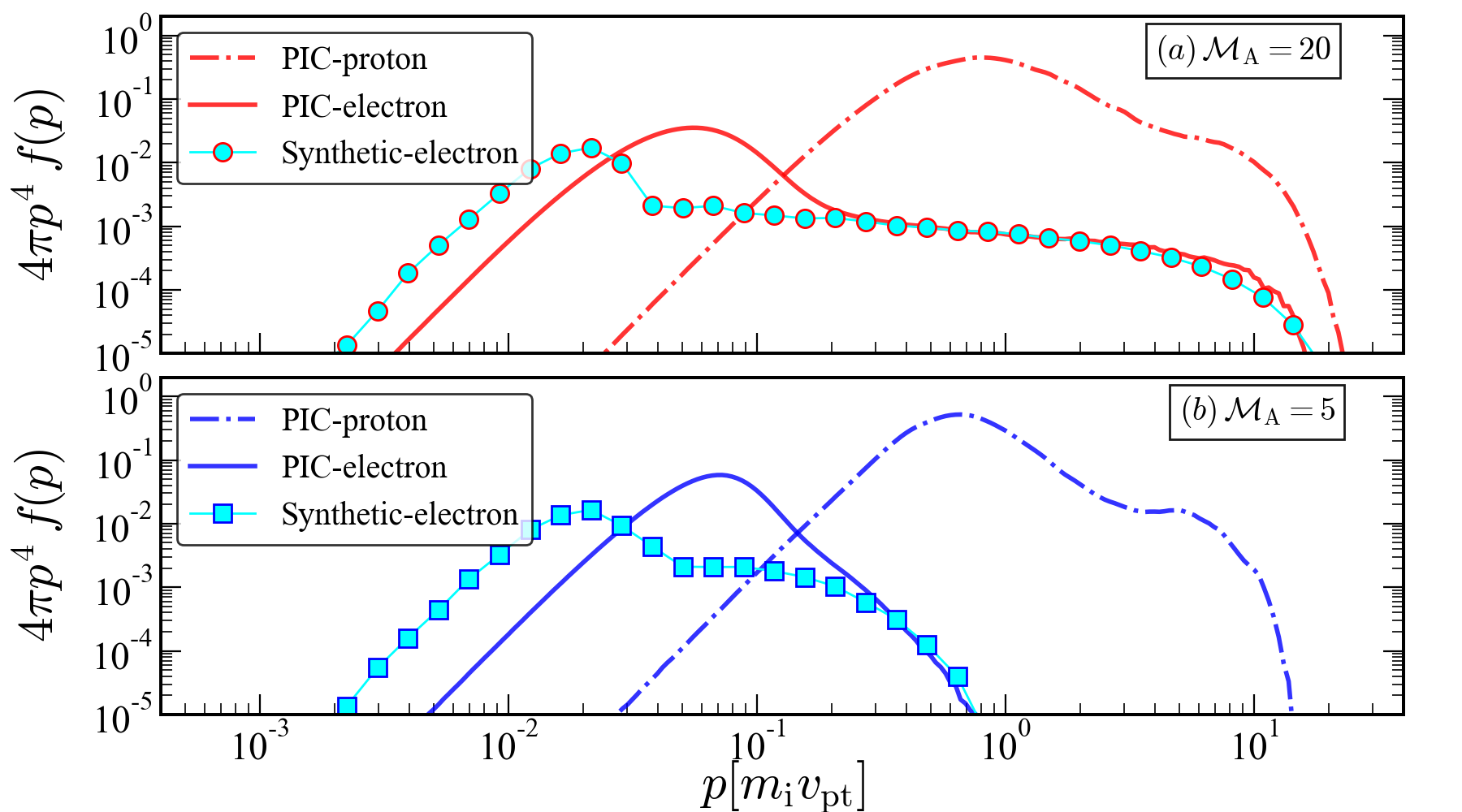}
    \caption{
 Comparison of electron spectra between PIC simulations (solid curves, measured at $275\,\omega_{\rm ci}^{-1}$ in the shock downstream) and synthetic model (cyan-filled symbols).
 }
 \label{fig:synthetic}
\end{figure}

Fig.~\ref{fig:synthetic} shows that the synthetic electron spectra obtained with this procedure (circles/squares) agree well with our PIC simulations for NT electrons.
They fail to explain the thermal peaks because we do not explicitly include any heating mechanism for the thermal electrons transmitted downstream.
As kinetic simulations (and observations) indicate that downstream thermal electrons are nearly in equipartition with protons \citep[][]{park+15, raymond+23, vanthieghem+24}, the NT electron tail emerges only above the downstream thermal distribution, i.e., when their momenta exceed
$p_{\rm th,e}\approx 3 \sqrt{m_{\rm e}/m_{\rm i}}\, m_{\rm i}v_{\rm pt}$.
Thus, the measurable fraction of NT electrons in the downstream is $\eta_{\rm e}\equiv \eta_{\rm inj}(p\geq  p_{\rm th,e})\sim 0.002 (\eta_{\rm inj}/0.05)/\sqrt{m_{\rm R}/100}$, as discussed in our PIC survey paper \citep[][]{gupta+24b}. 

Our minimal model can be used to explain the ratio of the normalization of the spectra of NT electrons and protons, $K_{\rm ep}$, inferred in Galactic CRs and CR sources \citep[e.g.,][]{morlino+10,morlino+12,merten+17, battiston20}. 
Because the NT electron tail begins at a lower momentum than that of NT protons (dash-dotted curve in Fig.~\ref{fig:synthetic}), electrons must undergo more DSA cycles to reach the same momentum as NT protons: 
since at each cycle a sizable fraction of the electrons is lost, $K_{\rm ep}$ remains well below unity \citep[][]{park+15,arbutina+21,gupta+24b, shalaby+22}. 
$K_{\rm ep}$ can be estimated by plugging the NT electron injection fraction $\eta_{\rm e}\sim 10^{-3}$ found in our minimal model and the proton injection fraction $\eta_{\rm i}\sim 10^{-2}$ reported in \citep[][]{caprioli+15} into the following expression 
\begin{equation}
    K_{\rm ep}\equiv \frac{f_{\rm e}(p)}{f_{\rm i}(p)}\approx  \frac{\eta_{\rm e}}{\eta_{\rm i}}\, \left(\frac{p}{p_{\rm inj}}\right)^{q_{\rm i}-q_{\rm e}} \frac{1}{m_{\rm R}^{(q_{\rm e}-3)/2}},
\end{equation}
which returns $K_{\rm ep}\sim 10^{-3}$ for the spectral indices $q_{\rm e}\approx q_{\rm i}= 4$, and $m_{\rm i}/m_{\rm e}=1836$.

Our findings are consistent with previous simulations of both quasi-parallel \citep[][]{park+15,crumley+19,arbutina+21,shalaby+22,gupta+24b} and quasi-perpendicular \citep{xu+20} shocks, where the electron DSA power-law begins at momenta lower than the proton $p_{\rm inj}$.
\section{Conclusions}\label{sec:summary}
%
We have studied the criteria of electron injection in DSA using PIC simulations of quasi-parallel shocks.
Our findings challenge the traditional assumption that the injection threshold is in momentum, demonstrating instead that speed is key. 

The main points of electron injection are:
1) electrons reflect off the shock via magnetic mirroring,
2) since mirroring depends on electron velocity, higher thermal speeds (i.e., low electron sonic Mach shocks) increase the likelihood of reflection, indicating that electron reflection at high-$\Ms$ shocks requires upstream preheating \citep[][]{gupta+24a}; 
otherwise, the fraction of particles with pitch angle conducive to reflection is negligible; 
3) electrons follow field lines and bounce between the shock and the upstream fluctuations, gaining energy as predicted by DSA (Fig.~\ref{fig:M20gain});
4) electrons need more DSA cycles to achieve momenta where both species are NT, which naturally leads to $K_{\rm ep}\ll 1$, in good agreement with observations \citep[][]{vanweeren+10,su+10,morlino+12,SN1006HESS,masters+16,liu+19,raptis+25}.

We also show that the rate of evolution of NT tails depends critically on the turbulence driven by accelerated protons.
In the absence of preexisting turbulence \citep[e.g.,][]{karol+23}, weak self-generated one may hinder preheating and prevent NT electrons from returning to the shock, suppressing further acceleration \citep[][]{gupta+24b}.

We leave to future works a full characterization of weak shocks and more oblique shocks, where transport is more complicated \citep[e.g.,][]{kirk+96, bell+11}, as well as a discussion of the potential role of non-linear magnetic fluctuations both in the upstream and in the downstream \citep[][]{haggerty+20, caprioli+20, zekovic+24}.\\

\begin{acknowledgments}
S.G. thanks Robert Ewart and Luca Orusa for insightful discussions.
We acknowledge the computational resources provided by the University of Chicago Research Computing Center and Princeton Research Computing.
D.C.~was partially supported by NASA through grants 80NSSC20K1273 and 80NSSC18K1218 and NSF through grants AST-1909778, PHY-2010240, and AST-2009326,
A.S.~acknowledges the support of grants from NSF (PHY-2206607) and the Simons Foundation (MP-SCMPS-00001470).
\end{acknowledgments}

\bibliography{Total}

\begin{thebibliography}{71}%
\makeatletter
\providecommand \@ifxundefined [1]{%
 \@ifx{#1\undefined}
}%
\providecommand \@ifnum [1]{%
 \ifnum #1\expandafter \@firstoftwo
 \else \expandafter \@secondoftwo
 \fi
}%
\providecommand \@ifx [1]{%
 \ifx #1\expandafter \@firstoftwo
 \else \expandafter \@secondoftwo
 \fi
}%
\providecommand \natexlab [1]{#1}%
\providecommand \enquote  [1]{``#1''}%
\providecommand \bibnamefont  [1]{#1}%
\providecommand \bibfnamefont [1]{#1}%
\providecommand \citenamefont [1]{#1}%
\providecommand \href@noop [0]{\@secondoftwo}%
\providecommand \href [0]{\begingroup \@sanitize@url \@href}%
\providecommand \@href[1]{\@@startlink{#1}\@@href}%
\providecommand \@@href[1]{\endgroup#1\@@endlink}%
\providecommand \@sanitize@url [0]{\catcode `\\12\catcode `\$12\catcode
  `\&12\catcode `\#12\catcode `\^12\catcode `\_12\catcode `\%12\relax}%
\providecommand \@@startlink[1]{}%
\providecommand \@@endlink[0]{}%
\providecommand \url  [0]{\begingroup\@sanitize@url \@url }%
\providecommand \@url [1]{\endgroup\@href {#1}{\urlprefix }}%
\providecommand \urlprefix  [0]{URL }%
\providecommand \Eprint [0]{\href }%
\providecommand \doibase [0]{https://doi.org/}%
\providecommand \selectlanguage [0]{\@gobble}%
\providecommand \bibinfo  [0]{\@secondoftwo}%
\providecommand \bibfield  [0]{\@secondoftwo}%
\providecommand \translation [1]{[#1]}%
\providecommand \BibitemOpen [0]{}%
\providecommand \bibitemStop [0]{}%
\providecommand \bibitemNoStop [0]{.\EOS\space}%
\providecommand \EOS [0]{\spacefactor3000\relax}%
\providecommand \BibitemShut  [1]{\csname bibitem#1\endcsname}%
\let\auto@bib@innerbib\@empty
\bibitem [{\citenamefont {Caprioli}(2015)}]{caprioli15p}%
  \BibitemOpen
  \bibfield  {author} {\bibinfo {author} {\bibfnamefont {D.}~\bibnamefont
  {Caprioli}},\ }\bibfield  {title} {\bibinfo {title} {Cosmic-ray acceleration
  and propagation},\ }in\ \href@noop {} {\emph {\bibinfo {booktitle} {34th
  International Cosmic Ray Conference (ICRC2015)}}},\ \bibinfo {series}
  {International Cosmic Ray Conference}, Vol.~\bibinfo {volume} {34},\ \bibinfo
  {editor} {edited by\ \bibinfo {editor} {\bibfnamefont {A.~S.}\ \bibnamefont
  {{Borisov}}}, \bibinfo {editor} {\bibfnamefont {V.~G.}\ \bibnamefont
  {{Denisova}}}, \bibinfo {editor} {\bibfnamefont {Z.~M.}\ \bibnamefont
  {{Guseva}}}, \bibinfo {editor} {\bibfnamefont {E.~A.}\ \bibnamefont
  {{Kanevskaya}}}, \bibinfo {editor} {\bibfnamefont {M.~G.}\ \bibnamefont
  {{Kogan}}}, \bibinfo {editor} {\bibfnamefont {A.~E.}\ \bibnamefont
  {{Morozov}}}, \bibinfo {editor} {\bibfnamefont {V.~S.}\ \bibnamefont
  {{Puchkov}}}, \bibinfo {editor} {\bibfnamefont {S.~E.}\ \bibnamefont
  {{Pyatovsky}}}, \bibinfo {editor} {\bibfnamefont {G.~P.}\ \bibnamefont
  {{Shoziyoev}}}, \bibinfo {editor} {\bibfnamefont {M.~D.}\ \bibnamefont
  {{Smirnova}}}, \bibinfo {editor} {\bibfnamefont {A.~V.}\ \bibnamefont
  {{Vargasov}}}, \bibinfo {editor} {\bibfnamefont {V.~I.}\ \bibnamefont
  {{Galkin}}}, \bibinfo {editor} {\bibfnamefont {S.~I.}\ \bibnamefont
  {{Nazarov}}},\ and\ \bibinfo {editor} {\bibfnamefont {R.~A.}\ \bibnamefont
  {{Mukhamedshin}}}}\ (\bibinfo {year} {2015})\ p.~\bibinfo {pages} {8},\
  \Eprint {https://arxiv.org/abs/1510.07042} {arXiv:1510.07042 [astro-ph.HE]}
  \BibitemShut {NoStop}%
\bibitem [{\citenamefont {{Marcowith}}\ \emph {et~al.}(2016)\citenamefont
  {{Marcowith}}, \citenamefont {{Bret}}, \citenamefont {{Bykov}}, \citenamefont
  {{Dieckman}}, \citenamefont {{O'C Drury}}, \citenamefont {{Lemb{\`e}ge}},
  \citenamefont {{Lemoine}}, \citenamefont {{Morlino}}, \citenamefont
  {{Murphy}}, \citenamefont {{Pelletier}}, \citenamefont {{Plotnikov}},
  \citenamefont {{Reville}}, \citenamefont {{Riquelme}}, \citenamefont
  {{Sironi}},\ and\ \citenamefont {{Stockem Novo}}}]{marcowith+16}%
  \BibitemOpen
  \bibfield  {author} {\bibinfo {author} {\bibfnamefont {A.}~\bibnamefont
  {{Marcowith}}}, \bibinfo {author} {\bibfnamefont {A.}~\bibnamefont {{Bret}}},
  \bibinfo {author} {\bibfnamefont {A.}~\bibnamefont {{Bykov}}}, \bibinfo
  {author} {\bibfnamefont {M.~E.}\ \bibnamefont {{Dieckman}}}, \bibinfo
  {author} {\bibfnamefont {L.}~\bibnamefont {{O'C Drury}}}, \bibinfo {author}
  {\bibfnamefont {B.}~\bibnamefont {{Lemb{\`e}ge}}}, \bibinfo {author}
  {\bibfnamefont {M.}~\bibnamefont {{Lemoine}}}, \bibinfo {author}
  {\bibfnamefont {G.}~\bibnamefont {{Morlino}}}, \bibinfo {author}
  {\bibfnamefont {G.}~\bibnamefont {{Murphy}}}, \bibinfo {author}
  {\bibfnamefont {G.}~\bibnamefont {{Pelletier}}}, \bibinfo {author}
  {\bibfnamefont {I.}~\bibnamefont {{Plotnikov}}}, \bibinfo {author}
  {\bibfnamefont {B.}~\bibnamefont {{Reville}}}, \bibinfo {author}
  {\bibfnamefont {M.}~\bibnamefont {{Riquelme}}}, \bibinfo {author}
  {\bibfnamefont {L.}~\bibnamefont {{Sironi}}},\ and\ \bibinfo {author}
  {\bibfnamefont {A.}~\bibnamefont {{Stockem Novo}}},\ }\bibfield  {title}
  {\bibinfo {title} {{The microphysics of collisionless shock waves}},\ }\href
  {https://doi.org/10.1088/0034-4885/79/4/046901} {\bibfield  {journal}
  {\bibinfo  {journal} {Reports on Progress in Physics}\ }\textbf {\bibinfo
  {volume} {79}},\ \bibinfo {eid} {046901} (\bibinfo {year} {2016})},\ \Eprint
  {https://arxiv.org/abs/1604.00318} {arXiv:1604.00318 [astro-ph.HE]}
  \BibitemShut {NoStop}%
\bibitem [{\citenamefont {{Bohdan}}(2023)}]{bohdan22}%
  \BibitemOpen
  \bibfield  {author} {\bibinfo {author} {\bibfnamefont {A.}~\bibnamefont
  {{Bohdan}}},\ }\bibfield  {title} {\bibinfo {title} {{Electron acceleration
  in supernova remnants}},\ }\href {https://doi.org/10.1088/1361-6587/aca5b2}
  {\bibfield  {journal} {\bibinfo  {journal} {Plasma Physics and Controlled
  Fusion}\ }\textbf {\bibinfo {volume} {65}},\ \bibinfo {eid} {014002}
  (\bibinfo {year} {2023})},\ \Eprint {https://arxiv.org/abs/2211.13992}
  {arXiv:2211.13992 [astro-ph.HE]} \BibitemShut {NoStop}%
\bibitem [{\citenamefont {{Krymskii}}(1977)}]{krymskii77}%
  \BibitemOpen
  \bibfield  {author} {\bibinfo {author} {\bibfnamefont {G.~F.}\ \bibnamefont
  {{Krymskii}}},\ }\bibfield  {title} {\bibinfo {title} {{A regular mechanism
  for the acceleration of charged particles on the front of a shock wave}},\
  }\href {https://ui.adsabs.harvard.edu/abs/1977DoSSR.234R1306K} {\bibfield
  {journal} {\bibinfo  {journal} {Akademiia Nauk SSSR Doklady}\ }\textbf
  {\bibinfo {volume} {234}},\ \bibinfo {pages} {1306} (\bibinfo {year}
  {1977})}\BibitemShut {NoStop}%
\bibitem [{\citenamefont {{Axford}}\ \emph {et~al.}(1977)\citenamefont
  {{Axford}}, \citenamefont {{Leer}},\ and\ \citenamefont
  {{Skadron}}}]{axford+77p}%
  \BibitemOpen
  \bibfield  {author} {\bibinfo {author} {\bibfnamefont {W.~I.}\ \bibnamefont
  {{Axford}}}, \bibinfo {author} {\bibfnamefont {E.}~\bibnamefont {{Leer}}},\
  and\ \bibinfo {author} {\bibfnamefont {G.}~\bibnamefont {{Skadron}}},\
  }\bibfield  {title} {\bibinfo {title} {{Acceleration of Cosmic Rays at Shock
  Fronts (Abstract)}},\ }in\ \href
  {http://adsabs.harvard.edu/abs/1977ICRC....2..273A} {\emph {\bibinfo
  {booktitle} {\emph{Acceleration of Cosmic Rays at Shock Fronts}}}},\ \bibinfo
  {series} {International Cosmic Ray Conference}, Vol.~\bibinfo {volume} {2}\
  (\bibinfo {year} {1977})\ pp.\ \bibinfo {pages} {273--+}\BibitemShut
  {NoStop}%
\bibitem [{\citenamefont {{Blandford}}\ and\ \citenamefont
  {{Ostriker}}(1978)}]{blandford+78}%
  \BibitemOpen
  \bibfield  {author} {\bibinfo {author} {\bibfnamefont {R.~D.}\ \bibnamefont
  {{Blandford}}}\ and\ \bibinfo {author} {\bibfnamefont {J.~P.}\ \bibnamefont
  {{Ostriker}}},\ }\bibfield  {title} {\bibinfo {title} {{Particle acceleration
  by astrophysical shocks}},\ }\href {https://doi.org/10.1086/182658}
  {\bibfield  {journal} {\bibinfo  {journal} {ApJL}\ }\textbf {\bibinfo
  {volume} {221}},\ \bibinfo {pages} {L29} (\bibinfo {year}
  {1978})}\BibitemShut {NoStop}%
\bibitem [{\citenamefont {{Bell}}(1978)}]{bell78a}%
  \BibitemOpen
  \bibfield  {author} {\bibinfo {author} {\bibfnamefont {A.~R.}\ \bibnamefont
  {{Bell}}},\ }\bibfield  {title} {\bibinfo {title} {{The acceleration of
  cosmic rays in shock fronts. I}},\ }\href
  {https://ui.adsabs.harvard.edu/abs/1978MNRAS.182..147B/abstract} {\bibfield
  {journal} {\bibinfo  {journal} {MNRAS}\ }\textbf {\bibinfo {volume} {182}},\
  \bibinfo {pages} {147} (\bibinfo {year} {1978})}\BibitemShut {NoStop}%
\bibitem [{\citenamefont {{van Weeren}}\ \emph {et~al.}(2010)\citenamefont
  {{van Weeren}}, \citenamefont {{R{\"o}ttgering}}, \citenamefont
  {{Br{\"u}ggen}},\ and\ \citenamefont {{Hoeft}}}]{vanweeren+10}%
  \BibitemOpen
  \bibfield  {author} {\bibinfo {author} {\bibfnamefont {R.~J.}\ \bibnamefont
  {{van Weeren}}}, \bibinfo {author} {\bibfnamefont {H.~J.~A.}\ \bibnamefont
  {{R{\"o}ttgering}}}, \bibinfo {author} {\bibfnamefont {M.}~\bibnamefont
  {{Br{\"u}ggen}}},\ and\ \bibinfo {author} {\bibfnamefont {M.}~\bibnamefont
  {{Hoeft}}},\ }\bibfield  {title} {\bibinfo {title} {{Particle Acceleration on
  Megaparsec Scales in a Merging Galaxy Cluster}},\ }\href
  {https://doi.org/10.1126/science.1194293} {\bibfield  {journal} {\bibinfo
  {journal} {Science}\ }\textbf {\bibinfo {volume} {330}},\ \bibinfo {pages}
  {347} (\bibinfo {year} {2010})},\ \Eprint {https://arxiv.org/abs/1010.4306}
  {arXiv:1010.4306 [astro-ph.CO]} \BibitemShut {NoStop}%
\bibitem [{\citenamefont {{Su}}\ \emph {et~al.}(2010)\citenamefont {{Su}},
  \citenamefont {{Slatyer}},\ and\ \citenamefont {{Finkbeiner}}}]{su+10}%
  \BibitemOpen
  \bibfield  {author} {\bibinfo {author} {\bibfnamefont {M.}~\bibnamefont
  {{Su}}}, \bibinfo {author} {\bibfnamefont {T.~R.}\ \bibnamefont
  {{Slatyer}}},\ and\ \bibinfo {author} {\bibfnamefont {D.~P.}\ \bibnamefont
  {{Finkbeiner}}},\ }\bibfield  {title} {\bibinfo {title} {{Giant Gamma-ray
  Bubbles from Fermi-LAT: Active Galactic Nucleus Activity or Bipolar Galactic
  Wind?}},\ }\href {https://doi.org/10.1088/0004-637X/724/2/1044} {\bibfield
  {journal} {\bibinfo  {journal} {\apj}\ }\textbf {\bibinfo {volume} {724}},\
  \bibinfo {pages} {1044} (\bibinfo {year} {2010})},\ \Eprint
  {https://arxiv.org/abs/1005.5480} {arXiv:1005.5480 [astro-ph.HE]}
  \BibitemShut {NoStop}%
\bibitem [{\citenamefont {{Morlino}}\ and\ \citenamefont
  {{Caprioli}}(2012)}]{morlino+12}%
  \BibitemOpen
  \bibfield  {author} {\bibinfo {author} {\bibfnamefont {G.}~\bibnamefont
  {{Morlino}}}\ and\ \bibinfo {author} {\bibfnamefont {D.}~\bibnamefont
  {{Caprioli}}},\ }\bibfield  {title} {\bibinfo {title} {{Strong evidence for
  hadron acceleration in Tycho's supernova remnant}},\ }\href
  {https://doi.org/10.1051/0004-6361/201117855} {\bibfield  {journal} {\bibinfo
   {journal} {A\&A}\ }\textbf {\bibinfo {volume} {538}},\ \bibinfo {eid} {A81}
  (\bibinfo {year} {2012})},\ \Eprint {https://arxiv.org/abs/arXiv:1105.6342}
  {arXiv:arXiv:1105.6342 [astro-ph.HE]} \BibitemShut {NoStop}%
\bibitem [{\citenamefont {{Acero et al.}}(2010)}]{SN1006HESS}%
  \BibitemOpen
  \bibfield  {author} {\bibinfo {author} {\bibfnamefont {F.}~\bibnamefont
  {{Acero et al.}}},\ }\bibfield  {title} {\bibinfo {title} {{First detection
  of VHE {$\gamma$}-rays from SN 1006 by HESS}},\ }\href
  {https://doi.org/10.1051/0004-6361/200913916} {\bibfield  {journal} {\bibinfo
   {journal} {A\&A}\ }\textbf {\bibinfo {volume} {516}},\ \bibinfo {pages}
  {A62+} (\bibinfo {year} {2010})},\ \Eprint {https://arxiv.org/abs/1004.2124}
  {arXiv:1004.2124 [astro-ph.HE]} \BibitemShut {NoStop}%
\bibitem [{\citenamefont {Masters}\ \emph {et~al.}(2016)\citenamefont
  {Masters}, \citenamefont {Sulaiman}, \citenamefont {Sergis}, \citenamefont
  {Stawarz}, \citenamefont {Fujimoto}, \citenamefont {Coates},\ and\
  \citenamefont {Dougherty}}]{masters+16}%
  \BibitemOpen
  \bibfield  {author} {\bibinfo {author} {\bibfnamefont {A.}~\bibnamefont
  {Masters}}, \bibinfo {author} {\bibfnamefont {A.~H.}\ \bibnamefont
  {Sulaiman}}, \bibinfo {author} {\bibfnamefont {N.}~\bibnamefont {Sergis}},
  \bibinfo {author} {\bibfnamefont {L.}~\bibnamefont {Stawarz}}, \bibinfo
  {author} {\bibfnamefont {M.}~\bibnamefont {Fujimoto}}, \bibinfo {author}
  {\bibfnamefont {A.~J.}\ \bibnamefont {Coates}},\ and\ \bibinfo {author}
  {\bibfnamefont {M.~K.}\ \bibnamefont {Dougherty}},\ }\bibfield  {title}
  {\bibinfo {title} {Suprathermal electrons at saturn's bow shock},\ }\href
  {https://doi.org/10.3847/0004-637X/826/1/48} {\bibfield  {journal} {\bibinfo
  {journal} {The Astrophysical Journal}\ }\textbf {\bibinfo {volume} {826}},\
  \bibinfo {pages} {48} (\bibinfo {year} {2016})}\BibitemShut {NoStop}%
\bibitem [{\citenamefont {{Liu}}\ \emph {et~al.}(2019)\citenamefont {{Liu}},
  \citenamefont {{Angelopoulos}},\ and\ \citenamefont {{Lu}}}]{liu+19}%
  \BibitemOpen
  \bibfield  {author} {\bibinfo {author} {\bibfnamefont {T.~Z.}\ \bibnamefont
  {{Liu}}}, \bibinfo {author} {\bibfnamefont {V.}~\bibnamefont
  {{Angelopoulos}}},\ and\ \bibinfo {author} {\bibfnamefont {S.}~\bibnamefont
  {{Lu}}},\ }\bibfield  {title} {\bibinfo {title} {{Relativistic electrons
  generated at Earth's quasi-parallel bow shock}},\ }\href
  {https://doi.org/10.1126/sciadv.aaw1368} {\bibfield  {journal} {\bibinfo
  {journal} {Science Advances}\ }\textbf {\bibinfo {volume} {5}},\ \bibinfo
  {eid} {eaaw1368} (\bibinfo {year} {2019})}\BibitemShut {NoStop}%
\bibitem [{\citenamefont {{Jebaraj}}\ \emph {et~al.}(2024)\citenamefont
  {{Jebaraj}}, \citenamefont {{Agapitov}}, \citenamefont {{Krasnoselskikh}},
  \citenamefont {{Vuorinen}}, \citenamefont {{Gedalin}}, \citenamefont
  {{Choi}}, \citenamefont {{Palmerio}}, \citenamefont {{Wijsen}}, \citenamefont
  {{Dresing}}, \citenamefont {{Cohen}}, \citenamefont {{Kouloumvakos}},
  \citenamefont {{Balikhin}}, \citenamefont {{Vainio}}, \citenamefont
  {{Kilpua}}, \citenamefont {{Afanasiev}}, \citenamefont {{Verniero}},
  \citenamefont {{Mitchell}}, \citenamefont {{Trotta}}, \citenamefont {{Hill}},
  \citenamefont {{Raouafi}},\ and\ \citenamefont {{Bale}}}]{jebaraj+24}%
  \BibitemOpen
  \bibfield  {author} {\bibinfo {author} {\bibfnamefont {I.~C.}\ \bibnamefont
  {{Jebaraj}}}, \bibinfo {author} {\bibfnamefont {O.}~\bibnamefont
  {{Agapitov}}}, \bibinfo {author} {\bibfnamefont {V.}~\bibnamefont
  {{Krasnoselskikh}}}, \bibinfo {author} {\bibfnamefont {L.}~\bibnamefont
  {{Vuorinen}}}, \bibinfo {author} {\bibfnamefont {M.}~\bibnamefont
  {{Gedalin}}}, \bibinfo {author} {\bibfnamefont {K.-E.}\ \bibnamefont
  {{Choi}}}, \bibinfo {author} {\bibfnamefont {E.}~\bibnamefont {{Palmerio}}},
  \bibinfo {author} {\bibfnamefont {N.}~\bibnamefont {{Wijsen}}}, \bibinfo
  {author} {\bibfnamefont {N.}~\bibnamefont {{Dresing}}}, \bibinfo {author}
  {\bibfnamefont {C.}~\bibnamefont {{Cohen}}}, \bibinfo {author} {\bibfnamefont
  {A.}~\bibnamefont {{Kouloumvakos}}}, \bibinfo {author} {\bibfnamefont
  {M.}~\bibnamefont {{Balikhin}}}, \bibinfo {author} {\bibfnamefont
  {R.}~\bibnamefont {{Vainio}}}, \bibinfo {author} {\bibfnamefont
  {E.}~\bibnamefont {{Kilpua}}}, \bibinfo {author} {\bibfnamefont
  {A.}~\bibnamefont {{Afanasiev}}}, \bibinfo {author} {\bibfnamefont
  {J.}~\bibnamefont {{Verniero}}}, \bibinfo {author} {\bibfnamefont {J.~G.}\
  \bibnamefont {{Mitchell}}}, \bibinfo {author} {\bibfnamefont
  {D.}~\bibnamefont {{Trotta}}}, \bibinfo {author} {\bibfnamefont
  {M.}~\bibnamefont {{Hill}}}, \bibinfo {author} {\bibfnamefont
  {N.}~\bibnamefont {{Raouafi}}},\ and\ \bibinfo {author} {\bibfnamefont
  {S.~D.}\ \bibnamefont {{Bale}}},\ }\bibfield  {title} {\bibinfo {title}
  {{Acceleration of Electrons and Ions by an ``Almost'' Astrophysical Shock in
  the Heliosphere}},\ }\href {https://doi.org/10.3847/2041-8213/ad4daa}
  {\bibfield  {journal} {\bibinfo  {journal} {\apjl}\ }\textbf {\bibinfo
  {volume} {968}},\ \bibinfo {eid} {L8} (\bibinfo {year} {2024})},\ \Eprint
  {https://arxiv.org/abs/2405.07074} {arXiv:2405.07074 [physics.space-ph]}
  \BibitemShut {NoStop}%
\bibitem [{\citenamefont {{Giuffrida}}\ \emph {et~al.}(2022)\citenamefont
  {{Giuffrida}}, \citenamefont {{Miceli}}, \citenamefont {{Caprioli}},
  \citenamefont {{Decourchelle}}, \citenamefont {{Vink}}, \citenamefont
  {{Orlando}}, \citenamefont {{Bocchino}}, \citenamefont {{Greco}},\ and\
  \citenamefont {{Peres}}}]{giuffrida+22}%
  \BibitemOpen
  \bibfield  {author} {\bibinfo {author} {\bibfnamefont {R.}~\bibnamefont
  {{Giuffrida}}}, \bibinfo {author} {\bibfnamefont {M.}~\bibnamefont
  {{Miceli}}}, \bibinfo {author} {\bibfnamefont {D.}~\bibnamefont
  {{Caprioli}}}, \bibinfo {author} {\bibfnamefont {A.}~\bibnamefont
  {{Decourchelle}}}, \bibinfo {author} {\bibfnamefont {J.}~\bibnamefont
  {{Vink}}}, \bibinfo {author} {\bibfnamefont {S.}~\bibnamefont {{Orlando}}},
  \bibinfo {author} {\bibfnamefont {F.}~\bibnamefont {{Bocchino}}}, \bibinfo
  {author} {\bibfnamefont {E.}~\bibnamefont {{Greco}}},\ and\ \bibinfo {author}
  {\bibfnamefont {G.}~\bibnamefont {{Peres}}},\ }\bibfield  {title} {\bibinfo
  {title} {{The supernova remnant SN 1006 as a Galactic particle
  accelerator}},\ }\href {https://doi.org/10.1038/s41467-022-32781-4}
  {\bibfield  {journal} {\bibinfo  {journal} {Nat. Comm.}\ }\textbf {\bibinfo
  {volume} {13}},\ \bibinfo {eid} {5098} (\bibinfo {year} {2022})},\ \Eprint
  {https://arxiv.org/abs/2208.14491} {arXiv:2208.14491 [astro-ph.HE]}
  \BibitemShut {NoStop}%
\bibitem [{\citenamefont {{Blasi}}\ \emph {et~al.}(2005)\citenamefont
  {{Blasi}}, \citenamefont {{Gabici}},\ and\ \citenamefont
  {{Vannoni}}}]{blasi+05}%
  \BibitemOpen
  \bibfield  {author} {\bibinfo {author} {\bibfnamefont {P.}~\bibnamefont
  {{Blasi}}}, \bibinfo {author} {\bibfnamefont {S.}~\bibnamefont {{Gabici}}},\
  and\ \bibinfo {author} {\bibfnamefont {G.}~\bibnamefont {{Vannoni}}},\
  }\bibfield  {title} {\bibinfo {title} {{On the role of injection in kinetic
  approaches to non-linear particle acceleration at non-relativistic shock
  waves}},\ }\href {https://doi.org/10.1111/j.1365-2966.2005.09227.x}
  {\bibfield  {journal} {\bibinfo  {journal} {MNRAS}\ }\textbf {\bibinfo
  {volume} {361}},\ \bibinfo {pages} {907} (\bibinfo {year} {2005})},\ \Eprint
  {https://arxiv.org/abs/astro-ph/0505351} {astro-ph/0505351} \BibitemShut
  {NoStop}%
\bibitem [{\citenamefont {{Malkov}}\ and\ \citenamefont {{O'C.
  Drury}}(2001)}]{malkov+01}%
  \BibitemOpen
  \bibfield  {author} {\bibinfo {author} {\bibfnamefont {M.~A.}\ \bibnamefont
  {{Malkov}}}\ and\ \bibinfo {author} {\bibfnamefont {L.}~\bibnamefont {{O'C.
  Drury}}},\ }\bibfield  {title} {\bibinfo {title} {{Nonlinear theory of
  diffusive acceleration of particles by shock waves}},\ }\href
  {http://adsabs.harvard.edu/abs/2001RPPh...64..429M} {\bibfield  {journal}
  {\bibinfo  {journal} {Rep. Prog. Phys.}\ }\textbf {\bibinfo {volume} {64}},\
  \bibinfo {pages} {429} (\bibinfo {year} {2001})}\BibitemShut {NoStop}%
\bibitem [{\citenamefont {{Kang}}\ \emph {et~al.}(2002)\citenamefont {{Kang}},
  \citenamefont {{Jones}},\ and\ \citenamefont {{Gieseler}}}]{kang+02}%
  \BibitemOpen
  \bibfield  {author} {\bibinfo {author} {\bibfnamefont {H.}~\bibnamefont
  {{Kang}}}, \bibinfo {author} {\bibfnamefont {T.~W.}\ \bibnamefont
  {{Jones}}},\ and\ \bibinfo {author} {\bibfnamefont {U.~D.~J.}\ \bibnamefont
  {{Gieseler}}},\ }\bibfield  {title} {\bibinfo {title} {{Numerical Studies of
  Cosmic-Ray Injection and Acceleration}},\ }\href
  {https://doi.org/10.1086/342724} {\bibfield  {journal} {\bibinfo  {journal}
  {\apj}\ }\textbf {\bibinfo {volume} {579}},\ \bibinfo {pages} {337} (\bibinfo
  {year} {2002})},\ \Eprint {https://arxiv.org/abs/astro-ph/0207410}
  {arXiv:astro-ph/0207410 [astro-ph]} \BibitemShut {NoStop}%
\bibitem [{\citenamefont {{Treumann}}(2009)}]{treumann09}%
  \BibitemOpen
  \bibfield  {author} {\bibinfo {author} {\bibfnamefont {R.~A.}\ \bibnamefont
  {{Treumann}}},\ }\bibfield  {title} {\bibinfo {title} {{Fundamentals of
  collisionless shocks for astrophysical application, 1. Non-relativistic
  shocks}},\ }\href {https://doi.org/10.1007/s00159-009-0024-2} {\bibfield
  {journal} {\bibinfo  {journal} {\aapr}\ }\textbf {\bibinfo {volume} {17}},\
  \bibinfo {pages} {409} (\bibinfo {year} {2009})}\BibitemShut {NoStop}%
\bibitem [{\citenamefont {{Caprioli}}\ \emph {et~al.}(2015)\citenamefont
  {{Caprioli}}, \citenamefont {{Pop}},\ and\ \citenamefont
  {{Spitkovsky}}}]{caprioli+15}%
  \BibitemOpen
  \bibfield  {author} {\bibinfo {author} {\bibfnamefont {D.}~\bibnamefont
  {{Caprioli}}}, \bibinfo {author} {\bibfnamefont {A.}~\bibnamefont {{Pop}}},\
  and\ \bibinfo {author} {\bibfnamefont {A.}~\bibnamefont {{Spitkovsky}}},\
  }\bibfield  {title} {\bibinfo {title} {{Simulations and Theory of Ion
  Injection at Non-relativistic Collisionless Shocks}},\ }\href@noop {}
  {\bibfield  {journal} {\bibinfo  {journal} {\apjl}\ }\textbf {\bibinfo
  {volume} {798}},\ \bibinfo {eid} {28} (\bibinfo {year} {2015})},\ \Eprint
  {https://arxiv.org/abs/1409.8291} {arXiv:1409.8291 [astro-ph.HE]}
  \BibitemShut {NoStop}%
\bibitem [{\citenamefont {{Caprioli}}\ and\ \citenamefont
  {{Spitkovsky}}(2014{\natexlab{a}})}]{caprioli+14a}%
  \BibitemOpen
  \bibfield  {author} {\bibinfo {author} {\bibfnamefont {D.}~\bibnamefont
  {{Caprioli}}}\ and\ \bibinfo {author} {\bibfnamefont {A.}~\bibnamefont
  {{Spitkovsky}}},\ }\bibfield  {title} {\bibinfo {title} {{Simulations of Ion
  Acceleration at Non-relativistic Shocks: I. Acceleration Efficiency}},\
  }\href {https://doi.org/10.1088/0004-637X/783/2/91} {\bibfield  {journal}
  {\bibinfo  {journal} {\apj}\ }\textbf {\bibinfo {volume} {783}},\ \bibinfo
  {eid} {91} (\bibinfo {year} {2014}{\natexlab{a}})},\ \Eprint
  {https://arxiv.org/abs/1310.2943} {arXiv:1310.2943 [astro-ph.HE]}
  \BibitemShut {NoStop}%
\bibitem [{\citenamefont {{Park}}\ \emph {et~al.}(2015)\citenamefont {{Park}},
  \citenamefont {{Caprioli}},\ and\ \citenamefont {{Spitkovsky}}}]{park+15}%
  \BibitemOpen
  \bibfield  {author} {\bibinfo {author} {\bibfnamefont {J.}~\bibnamefont
  {{Park}}}, \bibinfo {author} {\bibfnamefont {D.}~\bibnamefont {{Caprioli}}},\
  and\ \bibinfo {author} {\bibfnamefont {A.}~\bibnamefont {{Spitkovsky}}},\
  }\bibfield  {title} {\bibinfo {title} {{Simultaneous Acceleration of Protons
  and Electrons at Nonrelativistic Quasiparallel Collisionless Shocks}},\
  }\href {https://doi.org/10.1103/PhysRevLett.114.085003} {\bibfield  {journal}
  {\bibinfo  {journal} {Physical Review Letters}\ }\textbf {\bibinfo {volume}
  {114}},\ \bibinfo {eid} {085003} (\bibinfo {year} {2015})},\ \Eprint
  {https://arxiv.org/abs/1412.0672} {arXiv:1412.0672 [astro-ph.HE]}
  \BibitemShut {NoStop}%
\bibitem [{\citenamefont {{Kumar}}\ and\ \citenamefont
  {{Reville}}(2021)}]{kumar+21}%
  \BibitemOpen
  \bibfield  {author} {\bibinfo {author} {\bibfnamefont {N.}~\bibnamefont
  {{Kumar}}}\ and\ \bibinfo {author} {\bibfnamefont {B.}~\bibnamefont
  {{Reville}}},\ }\bibfield  {title} {\bibinfo {title} {{Nonthermal Particle
  Acceleration at Highly Oblique Nonrelativistic Shocks}},\ }\href
  {https://doi.org/10.3847/2041-8213/ac30e0} {\bibfield  {journal} {\bibinfo
  {journal} {\apjl}\ }\textbf {\bibinfo {volume} {921}},\ \bibinfo {eid} {L14}
  (\bibinfo {year} {2021})},\ \Eprint {https://arxiv.org/abs/2110.09939}
  {arXiv:2110.09939 [physics.plasm-ph]} \BibitemShut {NoStop}%
\bibitem [{\citenamefont {{Amano}}\ \emph {et~al.}(2022)\citenamefont
  {{Amano}}, \citenamefont {{Matsumoto}}, \citenamefont {{Bohdan}},
  \citenamefont {{Kobzar}}, \citenamefont {{Matsukiyo}}, \citenamefont {{Oka}},
  \citenamefont {{Niemiec}}, \citenamefont {{Pohl}},\ and\ \citenamefont
  {{Hoshino}}}]{amano+22}%
  \BibitemOpen
  \bibfield  {author} {\bibinfo {author} {\bibfnamefont {T.}~\bibnamefont
  {{Amano}}}, \bibinfo {author} {\bibfnamefont {Y.}~\bibnamefont
  {{Matsumoto}}}, \bibinfo {author} {\bibfnamefont {A.}~\bibnamefont
  {{Bohdan}}}, \bibinfo {author} {\bibfnamefont {O.}~\bibnamefont {{Kobzar}}},
  \bibinfo {author} {\bibfnamefont {S.}~\bibnamefont {{Matsukiyo}}}, \bibinfo
  {author} {\bibfnamefont {M.}~\bibnamefont {{Oka}}}, \bibinfo {author}
  {\bibfnamefont {J.}~\bibnamefont {{Niemiec}}}, \bibinfo {author}
  {\bibfnamefont {M.}~\bibnamefont {{Pohl}}},\ and\ \bibinfo {author}
  {\bibfnamefont {M.}~\bibnamefont {{Hoshino}}},\ }\bibfield  {title} {\bibinfo
  {title} {{Nonthermal electron acceleration at collisionless
  quasi-perpendicular shocks}},\ }\href
  {https://doi.org/10.1007/s41614-022-00093-1} {\bibfield  {journal} {\bibinfo
  {journal} {Reviews of Modern Plasma Physics}\ }\textbf {\bibinfo {volume}
  {6}},\ \bibinfo {eid} {29} (\bibinfo {year} {2022})},\ \Eprint
  {https://arxiv.org/abs/2209.03521} {arXiv:2209.03521 [astro-ph.HE]}
  \BibitemShut {NoStop}%
\bibitem [{\citenamefont {Ball}\ and\ \citenamefont {Melrose}(2001)}]{ball+01}%
  \BibitemOpen
  \bibfield  {author} {\bibinfo {author} {\bibfnamefont {L.}~\bibnamefont
  {Ball}}\ and\ \bibinfo {author} {\bibfnamefont {D.~B.}\ \bibnamefont
  {Melrose}},\ }\bibfield  {title} {\bibinfo {title} {Shock drift acceleration
  of electrons},\ }\href {https://doi.org/10.1071/AS01047} {\bibfield
  {journal} {\bibinfo  {journal} {PASA}\ }\textbf {\bibinfo {volume} {18}},\
  \bibinfo {pages} {361} (\bibinfo {year} {2001})}\BibitemShut {NoStop}%
\bibitem [{\citenamefont {{Hoshino}}\ and\ \citenamefont
  {{Shimada}}(2002)}]{hoshino+02}%
  \BibitemOpen
  \bibfield  {author} {\bibinfo {author} {\bibfnamefont {M.}~\bibnamefont
  {{Hoshino}}}\ and\ \bibinfo {author} {\bibfnamefont {N.}~\bibnamefont
  {{Shimada}}},\ }\bibfield  {title} {\bibinfo {title} {{Nonthermal Electrons
  at High Mach Number Shocks: Electron Shock Surfing Acceleration}},\ }\href
  {https://doi.org/10.1086/340454} {\bibfield  {journal} {\bibinfo  {journal}
  {\apj}\ }\textbf {\bibinfo {volume} {572}},\ \bibinfo {pages} {880} (\bibinfo
  {year} {2002})},\ \Eprint {https://arxiv.org/abs/astro-ph/0203073}
  {arXiv:astro-ph/0203073 [astro-ph]} \BibitemShut {NoStop}%
\bibitem [{\citenamefont {{Amano}}\ and\ \citenamefont
  {{Hoshino}}(2007)}]{amano+07}%
  \BibitemOpen
  \bibfield  {author} {\bibinfo {author} {\bibfnamefont {T.}~\bibnamefont
  {{Amano}}}\ and\ \bibinfo {author} {\bibfnamefont {M.}~\bibnamefont
  {{Hoshino}}},\ }\bibfield  {title} {\bibinfo {title} {{Electron Injection at
  High Mach Number Quasi-perpendicular Shocks: Surfing and Drift
  Acceleration}},\ }\href {https://doi.org/10.1086/513599} {\bibfield
  {journal} {\bibinfo  {journal} {\apj}\ }\textbf {\bibinfo {volume} {661}},\
  \bibinfo {pages} {190} (\bibinfo {year} {2007})},\ \Eprint
  {https://arxiv.org/abs/arXiv:astro-ph/0612204} {arXiv:astro-ph/0612204}
  \BibitemShut {NoStop}%
\bibitem [{\citenamefont {{Riquelme}}\ and\ \citenamefont
  {{Spitkovsky}}(2011)}]{riquelme+11}%
  \BibitemOpen
  \bibfield  {author} {\bibinfo {author} {\bibfnamefont {M.~A.}\ \bibnamefont
  {{Riquelme}}}\ and\ \bibinfo {author} {\bibfnamefont {A.}~\bibnamefont
  {{Spitkovsky}}},\ }\bibfield  {title} {\bibinfo {title} {{Electron Injection
  by Whistler Waves in Non-relativistic Shocks}},\ }\href
  {https://doi.org/10.1088/0004-637X/733/1/63} {\bibfield  {journal} {\bibinfo
  {journal} {\apj}\ }\textbf {\bibinfo {volume} {733}},\ \bibinfo {eid} {63}
  (\bibinfo {year} {2011})},\ \Eprint {https://arxiv.org/abs/1009.3319}
  {arXiv:1009.3319 [astro-ph.HE]} \BibitemShut {NoStop}%
\bibitem [{\citenamefont {{Matsumoto}}\ \emph {et~al.}(2012)\citenamefont
  {{Matsumoto}}, \citenamefont {{Amano}},\ and\ \citenamefont
  {{Hoshino}}}]{matsumoto+12}%
  \BibitemOpen
  \bibfield  {author} {\bibinfo {author} {\bibfnamefont {Y.}~\bibnamefont
  {{Matsumoto}}}, \bibinfo {author} {\bibfnamefont {T.}~\bibnamefont
  {{Amano}}},\ and\ \bibinfo {author} {\bibfnamefont {M.}~\bibnamefont
  {{Hoshino}}},\ }\bibfield  {title} {\bibinfo {title} {{Electron Accelerations
  at High Mach Number Shocks: Two-dimensional Particle-in-cell Simulations in
  Various Parameter Regimes}},\ }\href
  {https://doi.org/10.1088/0004-637X/755/2/109} {\bibfield  {journal} {\bibinfo
   {journal} {\apj}\ }\textbf {\bibinfo {volume} {755}},\ \bibinfo {eid} {109}
  (\bibinfo {year} {2012})},\ \Eprint {https://arxiv.org/abs/1204.6312}
  {arXiv:1204.6312 [astro-ph.HE]} \BibitemShut {NoStop}%
\bibitem [{\citenamefont {Guo}\ \emph {et~al.}(2014)\citenamefont {Guo},
  \citenamefont {Sironi},\ and\ \citenamefont {Narayan}}]{guo+14a}%
  \BibitemOpen
  \bibfield  {author} {\bibinfo {author} {\bibfnamefont {X.}~\bibnamefont
  {Guo}}, \bibinfo {author} {\bibfnamefont {L.}~\bibnamefont {Sironi}},\ and\
  \bibinfo {author} {\bibfnamefont {R.}~\bibnamefont {Narayan}},\ }\bibfield
  {title} {\bibinfo {title} {Non-thermal electron acceleration in low mach
  number collisionless shocks. i. particle energy spectra and acceleration
  mechanism},\ }\href {https://doi.org/10.1088/0004-637X/794/2/153} {\bibfield
  {journal} {\bibinfo  {journal} {\apj}\ }\textbf {\bibinfo {volume} {794}},\
  \bibinfo {eid} {153} (\bibinfo {year} {2014})},\ \Eprint
  {https://arxiv.org/abs/1406.5190} {arXiv:1406.5190 [astro-ph.HE]}
  \BibitemShut {NoStop}%
\bibitem [{\citenamefont {{Kato}}(2015)}]{kato15}%
  \BibitemOpen
  \bibfield  {author} {\bibinfo {author} {\bibfnamefont {T.~N.}\ \bibnamefont
  {{Kato}}},\ }\bibfield  {title} {\bibinfo {title} {{Particle Acceleration and
  Wave Excitation in Quasi-parallel High-Mach-number Collisionless Shocks:
  Particle-in-cell Simulation}},\ }\href
  {https://doi.org/10.1088/0004-637X/802/2/115} {\bibfield  {journal} {\bibinfo
   {journal} {\apj}\ }\textbf {\bibinfo {volume} {802}},\ \bibinfo {eid} {115}
  (\bibinfo {year} {2015})},\ \Eprint {https://arxiv.org/abs/1407.1971}
  {arXiv:1407.1971 [astro-ph.HE]} \BibitemShut {NoStop}%
\bibitem [{\citenamefont {Crumley}\ \emph {et~al.}(2019)\citenamefont
  {Crumley}, \citenamefont {Caprioli}, \citenamefont {Markoff},\ and\
  \citenamefont {Spitkovsky}}]{crumley+19}%
  \BibitemOpen
  \bibfield  {author} {\bibinfo {author} {\bibfnamefont {P.}~\bibnamefont
  {Crumley}}, \bibinfo {author} {\bibfnamefont {D.}~\bibnamefont {Caprioli}},
  \bibinfo {author} {\bibfnamefont {S.}~\bibnamefont {Markoff}},\ and\ \bibinfo
  {author} {\bibfnamefont {A.}~\bibnamefont {Spitkovsky}},\ }\bibfield  {title}
  {\bibinfo {title} {Kinetic simulations of mildly relativistic shocks:
  Particle acceleration in high mach number shocks},\ }\href
  {https://doi.org/10.1093/mnras/stz232} {\bibfield  {journal} {\bibinfo
  {journal} {\mnras}\ }\textbf {\bibinfo {volume} {485}},\ \bibinfo {pages}
  {5105} (\bibinfo {year} {2019})},\ \Eprint {https://arxiv.org/abs/1809.10809}
  {arXiv:1809.10809 [astro-ph.HE]} \BibitemShut {NoStop}%
\bibitem [{\citenamefont {{Bohdan}}\ \emph {et~al.}(2019)\citenamefont
  {{Bohdan}}, \citenamefont {{Niemiec}}, \citenamefont {{Pohl}}, \citenamefont
  {{Matsumoto}}, \citenamefont {{Amano}},\ and\ \citenamefont
  {{Hoshino}}}]{bohdan+19a}%
  \BibitemOpen
  \bibfield  {author} {\bibinfo {author} {\bibfnamefont {A.}~\bibnamefont
  {{Bohdan}}}, \bibinfo {author} {\bibfnamefont {J.}~\bibnamefont {{Niemiec}}},
  \bibinfo {author} {\bibfnamefont {M.}~\bibnamefont {{Pohl}}}, \bibinfo
  {author} {\bibfnamefont {Y.}~\bibnamefont {{Matsumoto}}}, \bibinfo {author}
  {\bibfnamefont {T.}~\bibnamefont {{Amano}}},\ and\ \bibinfo {author}
  {\bibfnamefont {M.}~\bibnamefont {{Hoshino}}},\ }\bibfield  {title} {\bibinfo
  {title} {{Kinetic Simulations of Nonrelativistic Perpendicular Shocks of
  Young Supernova Remnants. I. Electron Shock-surfing Acceleration}},\ }\href
  {https://doi.org/10.3847/1538-4357/ab1b6d} {\bibfield  {journal} {\bibinfo
  {journal} {\apj}\ }\textbf {\bibinfo {volume} {878}},\ \bibinfo {eid} {5}
  (\bibinfo {year} {2019})},\ \Eprint {https://arxiv.org/abs/1904.13153}
  {arXiv:1904.13153 [astro-ph.HE]} \BibitemShut {NoStop}%
\bibitem [{\citenamefont {{Xu}}\ \emph {et~al.}(2020)\citenamefont {{Xu}},
  \citenamefont {{Spitkovsky}},\ and\ \citenamefont {{Caprioli}}}]{xu+20}%
  \BibitemOpen
  \bibfield  {author} {\bibinfo {author} {\bibfnamefont {R.}~\bibnamefont
  {{Xu}}}, \bibinfo {author} {\bibfnamefont {A.}~\bibnamefont {{Spitkovsky}}},\
  and\ \bibinfo {author} {\bibfnamefont {D.}~\bibnamefont {{Caprioli}}},\
  }\bibfield  {title} {\bibinfo {title} {{Electron Acceleration in
  One-dimensional Nonrelativistic Quasi-perpendicular Collisionless Shocks}},\
  }\href {https://doi.org/10.3847/2041-8213/aba11e} {\bibfield  {journal}
  {\bibinfo  {journal} {\apjl}\ }\textbf {\bibinfo {volume} {897}},\ \bibinfo
  {eid} {L41} (\bibinfo {year} {2020})},\ \Eprint
  {https://arxiv.org/abs/1908.07890} {arXiv:1908.07890 [astro-ph.HE]}
  \BibitemShut {NoStop}%
\bibitem [{\citenamefont {{Arbutina}}\ and\ \citenamefont
  {{Zekovi{\'c}}}(2021)}]{arbutina+21}%
  \BibitemOpen
  \bibfield  {author} {\bibinfo {author} {\bibfnamefont {B.}~\bibnamefont
  {{Arbutina}}}\ and\ \bibinfo {author} {\bibfnamefont {V.}~\bibnamefont
  {{Zekovi{\'c}}}},\ }\bibfield  {title} {\bibinfo {title} {{Non-linear
  diffusive shock acceleration: A recipe for injection of electrons}},\ }\href
  {https://doi.org/10.1016/j.astropartphys.2020.102546} {\bibfield  {journal}
  {\bibinfo  {journal} {Astroparticle Physics}\ }\textbf {\bibinfo {volume}
  {127}},\ \bibinfo {eid} {102546} (\bibinfo {year} {2021})},\ \Eprint
  {https://arxiv.org/abs/2012.15117} {arXiv:2012.15117 [astro-ph.HE]}
  \BibitemShut {NoStop}%
\bibitem [{\citenamefont {{Shalaby}}\ \emph {et~al.}(2022)\citenamefont
  {{Shalaby}}, \citenamefont {{Lemmerz}}, \citenamefont {{Thomas}},\ and\
  \citenamefont {{Pfrommer}}}]{shalaby+22}%
  \BibitemOpen
  \bibfield  {author} {\bibinfo {author} {\bibfnamefont {M.}~\bibnamefont
  {{Shalaby}}}, \bibinfo {author} {\bibfnamefont {R.}~\bibnamefont
  {{Lemmerz}}}, \bibinfo {author} {\bibfnamefont {T.}~\bibnamefont
  {{Thomas}}},\ and\ \bibinfo {author} {\bibfnamefont {C.}~\bibnamefont
  {{Pfrommer}}},\ }\bibfield  {title} {\bibinfo {title} {{The Mechanism of
  Efficient Electron Acceleration at Parallel Nonrelativistic Shocks}},\ }\href
  {https://doi.org/10.3847/1538-4357/ac6ce7} {\bibfield  {journal} {\bibinfo
  {journal} {\apj}\ }\textbf {\bibinfo {volume} {932}},\ \bibinfo {eid} {86}
  (\bibinfo {year} {2022})},\ \Eprint {https://arxiv.org/abs/2202.05288}
  {arXiv:2202.05288 [astro-ph.HE]} \BibitemShut {NoStop}%
\bibitem [{\citenamefont {{Morris}}\ \emph {et~al.}(2022)\citenamefont
  {{Morris}}, \citenamefont {{Bohdan}}, \citenamefont {{Weidl}},\ and\
  \citenamefont {{Pohl}}}]{morris+22}%
  \BibitemOpen
  \bibfield  {author} {\bibinfo {author} {\bibfnamefont {P.~J.}\ \bibnamefont
  {{Morris}}}, \bibinfo {author} {\bibfnamefont {A.}~\bibnamefont {{Bohdan}}},
  \bibinfo {author} {\bibfnamefont {M.~S.}\ \bibnamefont {{Weidl}}},\ and\
  \bibinfo {author} {\bibfnamefont {M.}~\bibnamefont {{Pohl}}},\ }\bibfield
  {title} {\bibinfo {title} {{Preacceleration in the Electron Foreshock. I.
  Electron Acoustic Waves}},\ }\href {https://doi.org/10.3847/1538-4357/ac69c7}
  {\bibfield  {journal} {\bibinfo  {journal} {\apj}\ }\textbf {\bibinfo
  {volume} {931}},\ \bibinfo {eid} {129} (\bibinfo {year} {2022})},\ \Eprint
  {https://arxiv.org/abs/2204.11569} {arXiv:2204.11569 [physics.plasm-ph]}
  \BibitemShut {NoStop}%
\bibitem [{\citenamefont {{Zekovi{\'c}}}\ \emph {et~al.}(2024)\citenamefont
  {{Zekovi{\'c}}}, \citenamefont {{Spitkovsky}},\ and\ \citenamefont
  {{Hemler}}}]{zekovic+24}%
  \BibitemOpen
  \bibfield  {author} {\bibinfo {author} {\bibfnamefont {V.}~\bibnamefont
  {{Zekovi{\'c}}}}, \bibinfo {author} {\bibfnamefont {A.}~\bibnamefont
  {{Spitkovsky}}},\ and\ \bibinfo {author} {\bibfnamefont {Z.}~\bibnamefont
  {{Hemler}}},\ }\bibfield  {title} {\bibinfo {title} {{SLAMS-Propelled
  Electron Acceleration at High-Mach Number Astrophysical Shocks}},\ }\href
  {https://doi.org/10.48550/arXiv.2408.02084} {\bibfield  {journal} {\bibinfo
  {journal} {arXiv e-prints}\ ,\ \bibinfo {eid} {arXiv:2408.02084}} (\bibinfo
  {year} {2024})},\ \Eprint {https://arxiv.org/abs/2408.02084}
  {arXiv:2408.02084 [astro-ph.HE]} \BibitemShut {NoStop}%
\bibitem [{\citenamefont {{Caprioli}}(2015)}]{caprioli15}%
  \BibitemOpen
  \bibfield  {author} {\bibinfo {author} {\bibfnamefont {D.}~\bibnamefont
  {{Caprioli}}},\ }\bibfield  {title} {\bibinfo {title} {{''Espresso''
  Acceleration of Ultra-high-energy Cosmic Rays}},\ }\href
  {https://doi.org/10.1088/2041-8205/811/2/L38} {\bibfield  {journal} {\bibinfo
   {journal} {\apjl}\ }\textbf {\bibinfo {volume} {811}},\ \bibinfo {eid} {L38}
  (\bibinfo {year} {2015})},\ \Eprint {https://arxiv.org/abs/1505.06739}
  {arXiv:1505.06739 [astro-ph.HE]} \BibitemShut {NoStop}%
\bibitem [{\citenamefont {Brunetti}\ and\ \citenamefont
  {Jones}(2014)}]{brunetti+14}%
  \BibitemOpen
  \bibfield  {author} {\bibinfo {author} {\bibfnamefont {G.}~\bibnamefont
  {Brunetti}}\ and\ \bibinfo {author} {\bibfnamefont {T.~W.}\ \bibnamefont
  {Jones}},\ }\bibfield  {title} {\bibinfo {title} {Cosmic rays in galaxy
  clusters and their nonthermal emission},\ }\href
  {https://doi.org/10.1142/S0218271814300079} {\bibfield  {journal} {\bibinfo
  {journal} {International Journal of Modern Physics D}\ }\textbf {\bibinfo
  {volume} {23}},\ \bibinfo {eid} {1430007-98} (\bibinfo {year} {2014})},\
  \Eprint {https://arxiv.org/abs/1401.7519} {arXiv:1401.7519} \BibitemShut
  {NoStop}%
\bibitem [{\citenamefont {Diesing}\ \emph {et~al.}(2023)\citenamefont
  {Diesing}, \citenamefont {Metzger}, \citenamefont {Aydi}, \citenamefont
  {Chomiuk}, \citenamefont {Vurm}, \citenamefont {Gupta},\ and\ \citenamefont
  {Caprioli}}]{diesing+23}%
  \BibitemOpen
  \bibfield  {author} {\bibinfo {author} {\bibfnamefont {R.}~\bibnamefont
  {Diesing}}, \bibinfo {author} {\bibfnamefont {B.~D.}\ \bibnamefont
  {Metzger}}, \bibinfo {author} {\bibfnamefont {E.}~\bibnamefont {Aydi}},
  \bibinfo {author} {\bibfnamefont {L.}~\bibnamefont {Chomiuk}}, \bibinfo
  {author} {\bibfnamefont {I.}~\bibnamefont {Vurm}}, \bibinfo {author}
  {\bibfnamefont {S.}~\bibnamefont {Gupta}},\ and\ \bibinfo {author}
  {\bibfnamefont {D.}~\bibnamefont {Caprioli}},\ }\bibfield  {title} {\bibinfo
  {title} {Evidence for multiple shocks from the gamma-ray emission of rs
  ophiuchi},\ }\href {https://doi.org/10.3847/1538-4357/acc105} {\bibfield
  {journal} {\bibinfo  {journal} {\apj}\ }\textbf {\bibinfo {volume} {947}},\
  \bibinfo {pages} {70} (\bibinfo {year} {2023})}\BibitemShut {NoStop}%
\bibitem [{\citenamefont {Adriani}\ \emph {et~al.}(2011)\citenamefont
  {Adriani}, \citenamefont {Barbarino}, \citenamefont {Bazilevskaya},
  \citenamefont {Bellotti}, \citenamefont {Boezio}, \citenamefont {Bogomolov},
  \citenamefont {Bonechi}, \citenamefont {Bongi}, \citenamefont {Bonvicini},
  \citenamefont {Borisov}, \citenamefont {Bottai}, \citenamefont {Bruno},
  \citenamefont {Cafagna}, \citenamefont {Campana}, \citenamefont {Carbone},
  \citenamefont {Carlson}, \citenamefont {Casolino}, \citenamefont
  {Castellini}, \citenamefont {Consiglio}, \citenamefont {De~Pascale},
  \citenamefont {De~Santis}, \citenamefont {De~Simone}, \citenamefont
  {Di~Felice}, \citenamefont {Galper}, \citenamefont {Gillard}, \citenamefont
  {Grishantseva}, \citenamefont {Jerse}, \citenamefont {Karelin}, \citenamefont
  {Koldashov}, \citenamefont {Krutkov}, \citenamefont {Kvashnin}, \citenamefont
  {Leonov}, \citenamefont {Malakhov}, \citenamefont {Malvezzi}, \citenamefont
  {Marcelli}, \citenamefont {Mayorov}, \citenamefont {Menn}, \citenamefont
  {Mikhailov}, \citenamefont {Mocchiutti}, \citenamefont {Monaco},
  \citenamefont {Mori}, \citenamefont {Nikonov}, \citenamefont {Osteria},
  \citenamefont {Palma}, \citenamefont {Papini}, \citenamefont {Pearce},
  \citenamefont {Picozza}, \citenamefont {Pizzolotto}, \citenamefont {Ricci},
  \citenamefont {Ricciarini}, \citenamefont {Rossetto}, \citenamefont {Sarkar},
  \citenamefont {Simon}, \citenamefont {Sparvoli}, \citenamefont {Spillantini},
  \citenamefont {Stozhkov}, \citenamefont {Vacchi}, \citenamefont {Vannuccini},
  \citenamefont {Vasilyev}, \citenamefont {Voronov}, \citenamefont {Yurkin},
  \citenamefont {Wu}, \citenamefont {Zampa}, \citenamefont {Zampa},\ and\
  \citenamefont {Zverev}}]{pamela11}%
  \BibitemOpen
  \bibfield  {author} {\bibinfo {author} {\bibfnamefont {O.}~\bibnamefont
  {Adriani}}, \bibinfo {author} {\bibfnamefont {G.~C.}\ \bibnamefont
  {Barbarino}}, \bibinfo {author} {\bibfnamefont {G.~A.}\ \bibnamefont
  {Bazilevskaya}}, \bibinfo {author} {\bibfnamefont {R.}~\bibnamefont
  {Bellotti}}, \bibinfo {author} {\bibfnamefont {M.}~\bibnamefont {Boezio}},
  \bibinfo {author} {\bibfnamefont {E.~A.}\ \bibnamefont {Bogomolov}}, \bibinfo
  {author} {\bibfnamefont {L.}~\bibnamefont {Bonechi}}, \bibinfo {author}
  {\bibfnamefont {M.}~\bibnamefont {Bongi}}, \bibinfo {author} {\bibfnamefont
  {V.}~\bibnamefont {Bonvicini}}, \bibinfo {author} {\bibfnamefont
  {S.}~\bibnamefont {Borisov}}, \bibinfo {author} {\bibfnamefont
  {S.}~\bibnamefont {Bottai}}, \bibinfo {author} {\bibfnamefont
  {A.}~\bibnamefont {Bruno}}, \bibinfo {author} {\bibfnamefont
  {F.}~\bibnamefont {Cafagna}}, \bibinfo {author} {\bibfnamefont
  {D.}~\bibnamefont {Campana}}, \bibinfo {author} {\bibfnamefont
  {R.}~\bibnamefont {Carbone}}, \bibinfo {author} {\bibfnamefont
  {P.}~\bibnamefont {Carlson}}, \bibinfo {author} {\bibfnamefont
  {M.}~\bibnamefont {Casolino}}, \bibinfo {author} {\bibfnamefont
  {G.}~\bibnamefont {Castellini}}, \bibinfo {author} {\bibfnamefont
  {L.}~\bibnamefont {Consiglio}}, \bibinfo {author} {\bibfnamefont {M.~P.}\
  \bibnamefont {De~Pascale}}, \bibinfo {author} {\bibfnamefont
  {C.}~\bibnamefont {De~Santis}}, \bibinfo {author} {\bibfnamefont
  {N.}~\bibnamefont {De~Simone}}, \bibinfo {author} {\bibfnamefont
  {V.}~\bibnamefont {Di~Felice}}, \bibinfo {author} {\bibfnamefont {A.~M.}\
  \bibnamefont {Galper}}, \bibinfo {author} {\bibfnamefont {W.}~\bibnamefont
  {Gillard}}, \bibinfo {author} {\bibfnamefont {L.}~\bibnamefont
  {Grishantseva}}, \bibinfo {author} {\bibfnamefont {G.}~\bibnamefont {Jerse}},
  \bibinfo {author} {\bibfnamefont {A.~V.}\ \bibnamefont {Karelin}}, \bibinfo
  {author} {\bibfnamefont {S.~V.}\ \bibnamefont {Koldashov}}, \bibinfo {author}
  {\bibfnamefont {S.~Y.}\ \bibnamefont {Krutkov}}, \bibinfo {author}
  {\bibfnamefont {A.~N.}\ \bibnamefont {Kvashnin}}, \bibinfo {author}
  {\bibfnamefont {A.}~\bibnamefont {Leonov}}, \bibinfo {author} {\bibfnamefont
  {V.}~\bibnamefont {Malakhov}}, \bibinfo {author} {\bibfnamefont
  {V.}~\bibnamefont {Malvezzi}}, \bibinfo {author} {\bibfnamefont
  {L.}~\bibnamefont {Marcelli}}, \bibinfo {author} {\bibfnamefont {A.~G.}\
  \bibnamefont {Mayorov}}, \bibinfo {author} {\bibfnamefont {W.}~\bibnamefont
  {Menn}}, \bibinfo {author} {\bibfnamefont {V.~V.}\ \bibnamefont {Mikhailov}},
  \bibinfo {author} {\bibfnamefont {E.}~\bibnamefont {Mocchiutti}}, \bibinfo
  {author} {\bibfnamefont {A.}~\bibnamefont {Monaco}}, \bibinfo {author}
  {\bibfnamefont {N.}~\bibnamefont {Mori}}, \bibinfo {author} {\bibfnamefont
  {N.}~\bibnamefont {Nikonov}}, \bibinfo {author} {\bibfnamefont
  {G.}~\bibnamefont {Osteria}}, \bibinfo {author} {\bibfnamefont
  {F.}~\bibnamefont {Palma}}, \bibinfo {author} {\bibfnamefont
  {P.}~\bibnamefont {Papini}}, \bibinfo {author} {\bibfnamefont
  {M.}~\bibnamefont {Pearce}}, \bibinfo {author} {\bibfnamefont
  {P.}~\bibnamefont {Picozza}}, \bibinfo {author} {\bibfnamefont
  {C.}~\bibnamefont {Pizzolotto}}, \bibinfo {author} {\bibfnamefont
  {M.}~\bibnamefont {Ricci}}, \bibinfo {author} {\bibfnamefont {S.~B.}\
  \bibnamefont {Ricciarini}}, \bibinfo {author} {\bibfnamefont
  {L.}~\bibnamefont {Rossetto}}, \bibinfo {author} {\bibfnamefont
  {R.}~\bibnamefont {Sarkar}}, \bibinfo {author} {\bibfnamefont
  {M.}~\bibnamefont {Simon}}, \bibinfo {author} {\bibfnamefont
  {R.}~\bibnamefont {Sparvoli}}, \bibinfo {author} {\bibfnamefont
  {P.}~\bibnamefont {Spillantini}}, \bibinfo {author} {\bibfnamefont {Y.~I.}\
  \bibnamefont {Stozhkov}}, \bibinfo {author} {\bibfnamefont {A.}~\bibnamefont
  {Vacchi}}, \bibinfo {author} {\bibfnamefont {E.}~\bibnamefont {Vannuccini}},
  \bibinfo {author} {\bibfnamefont {G.}~\bibnamefont {Vasilyev}}, \bibinfo
  {author} {\bibfnamefont {S.~A.}\ \bibnamefont {Voronov}}, \bibinfo {author}
  {\bibfnamefont {Y.~T.}\ \bibnamefont {Yurkin}}, \bibinfo {author}
  {\bibfnamefont {J.}~\bibnamefont {Wu}}, \bibinfo {author} {\bibfnamefont
  {G.}~\bibnamefont {Zampa}}, \bibinfo {author} {\bibfnamefont
  {N.}~\bibnamefont {Zampa}},\ and\ \bibinfo {author} {\bibfnamefont {V.~G.}\
  \bibnamefont {Zverev}},\ }\bibfield  {title} {\bibinfo {title} {Pamela
  measurements of cosmic-ray proton and helium spectra},\ }\href
  {https://doi.org/10.1126/science.1199172} {\bibfield  {journal} {\bibinfo
  {journal} {Science}\ }\textbf {\bibinfo {volume} {332}},\ \bibinfo {pages}
  {69} (\bibinfo {year} {2011})},\ \Eprint
  {https://arxiv.org/abs/https://science.sciencemag.org/content/332/6025/69.full.pdf}
  {https://science.sciencemag.org/content/332/6025/69.full.pdf} \BibitemShut
  {NoStop}%
\bibitem [{\citenamefont {Aguilar et~al. [AMS~Collaboration]}(2019)}]{ams19b}%
  \BibitemOpen
  \bibfield  {author} {\bibinfo {author} {\bibfnamefont {M.}~\bibnamefont
  {Aguilar et~al. [AMS~Collaboration]}},\ }\bibfield  {title} {\bibinfo {title}
  {Towards understanding the origin of cosmic-ray electrons},\ }\href
  {https://doi.org/10.1103/PhysRevLett.122.101101} {\bibfield  {journal}
  {\bibinfo  {journal} {Phys. Rev. Lett.}\ }\textbf {\bibinfo {volume} {122}},\
  \bibinfo {pages} {101101} (\bibinfo {year} {2019})}\BibitemShut {NoStop}%
\bibitem [{\citenamefont {Bunemann}(1993)}]{buneman93}%
  \BibitemOpen
  \bibfield  {author} {\bibinfo {author} {\bibfnamefont {O.}~\bibnamefont
  {Bunemann}},\ }\bibinfo {title} {Computer space plasma physics:simulation
  techniques and software}\ (\bibinfo  {publisher} {Terra Scientific Publishing
  Company},\ \bibinfo {year} {1993})\ pp.\ \bibinfo {pages}
  {67--87}\BibitemShut {NoStop}%
\bibitem [{\citenamefont {{Spitkovsky}}(2005)}]{spitkovsky05}%
  \BibitemOpen
  \bibfield  {author} {\bibinfo {author} {\bibfnamefont {A.}~\bibnamefont
  {{Spitkovsky}}},\ }\bibfield  {title} {\bibinfo {title} {{Simulations of
  relativistic collisionless shocks: shock structure and particle
  acceleration}},\ }in\ \href {https://doi.org/10.1063/1.2141897} {\emph
  {\bibinfo {booktitle} {Astrophysical Sources of High Energy Particles and
  Radiation}}},\ \bibinfo {series} {American Institute of Physics Conference
  Series}, Vol.\ \bibinfo {volume} {801},\ \bibinfo {editor} {edited by\
  \bibinfo {editor} {\bibfnamefont {T.}~\bibnamefont {{Bulik}}}, \bibinfo
  {editor} {\bibfnamefont {B.}~\bibnamefont {{Rudak}}},\ and\ \bibinfo {editor}
  {\bibfnamefont {G.}~\bibnamefont {{Madejski}}}}\ (\bibinfo {year} {2005})\
  pp.\ \bibinfo {pages} {345--350},\ \Eprint
  {https://arxiv.org/abs/astro-ph/0603211} {astro-ph/0603211} \BibitemShut
  {NoStop}%
\bibitem [{\citenamefont {{Gupta}}\ \emph
  {et~al.}(2024{\natexlab{a}})\citenamefont {{Gupta}}, \citenamefont
  {{Caprioli}},\ and\ \citenamefont {{Spitkovsky}}}]{gupta+24b}%
  \BibitemOpen
  \bibfield  {author} {\bibinfo {author} {\bibfnamefont {S.}~\bibnamefont
  {{Gupta}}}, \bibinfo {author} {\bibfnamefont {D.}~\bibnamefont
  {{Caprioli}}},\ and\ \bibinfo {author} {\bibfnamefont {A.}~\bibnamefont
  {{Spitkovsky}}},\ }\bibfield  {title} {\bibinfo {title} {{Electron
  Acceleration at Quasi-parallel Nonrelativistic Shocks: A 1D Kinetic
  Survey}},\ }\href {https://doi.org/10.3847/1538-4357/ad7c4c} {\bibfield
  {journal} {\bibinfo  {journal} {\apj}\ }\textbf {\bibinfo {volume} {976}},\
  \bibinfo {eid} {10} (\bibinfo {year} {2024}{\natexlab{a}})},\ \Eprint
  {https://arxiv.org/abs/2408.16071} {arXiv:2408.16071 [astro-ph.HE]}
  \BibitemShut {NoStop}%
\bibitem [{\citenamefont {{Caprioli}}\ and\ \citenamefont
  {{Spitkovsky}}(2014{\natexlab{b}})}]{caprioli+14b}%
  \BibitemOpen
  \bibfield  {author} {\bibinfo {author} {\bibfnamefont {D.}~\bibnamefont
  {{Caprioli}}}\ and\ \bibinfo {author} {\bibfnamefont {A.}~\bibnamefont
  {{Spitkovsky}}},\ }\bibfield  {title} {\bibinfo {title} {{Simulations of Ion
  Acceleration at Non-relativistic Shocks: II. Magnetic Field Amplification}},\
  }\href {https://doi.org/10.1088/0004-637X/794/1/46} {\bibfield  {journal}
  {\bibinfo  {journal} {\apj}\ }\textbf {\bibinfo {volume} {794}},\ \bibinfo
  {eid} {46} (\bibinfo {year} {2014}{\natexlab{b}})},\ \Eprint
  {https://arxiv.org/abs/1401.7679} {arXiv:1401.7679 [astro-ph.HE]}
  \BibitemShut {NoStop}%
\bibitem [{\citenamefont {{Muschietti}}\ and\ \citenamefont
  {{Lemb{\`e}ge}}(2017)}]{muschietti+17}%
  \BibitemOpen
  \bibfield  {author} {\bibinfo {author} {\bibfnamefont {L.}~\bibnamefont
  {{Muschietti}}}\ and\ \bibinfo {author} {\bibfnamefont {B.}~\bibnamefont
  {{Lemb{\`e}ge}}},\ }\bibfield  {title} {\bibinfo {title} {{Two-stream
  instabilities from the lower-hybrid frequency to the electron cyclotron
  frequency: application to the front of quasi-perpendicular shocks}},\ }\href
  {https://doi.org/10.5194/angeo-35-1093-2017} {\bibfield  {journal} {\bibinfo
  {journal} {Annales Geophysicae}\ }\textbf {\bibinfo {volume} {35}},\ \bibinfo
  {pages} {1093} (\bibinfo {year} {2017})}\BibitemShut {NoStop}%
\bibitem [{\citenamefont {{Gupta}}\ \emph
  {et~al.}(2024{\natexlab{b}})\citenamefont {{Gupta}}, \citenamefont
  {{Caprioli}},\ and\ \citenamefont {{Spitkovsky}}}]{gupta+24a}%
  \BibitemOpen
  \bibfield  {author} {\bibinfo {author} {\bibfnamefont {S.}~\bibnamefont
  {{Gupta}}}, \bibinfo {author} {\bibfnamefont {D.}~\bibnamefont
  {{Caprioli}}},\ and\ \bibinfo {author} {\bibfnamefont {A.}~\bibnamefont
  {{Spitkovsky}}},\ }\bibfield  {title} {\bibinfo {title} {{Return Currents in
  Collisionless Shocks}},\ }\href {https://doi.org/10.3847/1538-4357/ad3e75}
  {\bibfield  {journal} {\bibinfo  {journal} {\apj}\ }\textbf {\bibinfo
  {volume} {968}},\ \bibinfo {eid} {17} (\bibinfo {year}
  {2024}{\natexlab{b}})},\ \Eprint {https://arxiv.org/abs/2312.13365}
  {arXiv:2312.13365 [astro-ph.HE]} \BibitemShut {NoStop}%
\bibitem [{\citenamefont {{Lichko}}\ \emph {et~al.}(2025)\citenamefont
  {{Lichko}}, \citenamefont {{Caprioli}}, \citenamefont {{Schroer}},\ and\
  \citenamefont {{Gupta}}}]{lichko+24}%
  \BibitemOpen
  \bibfield  {author} {\bibinfo {author} {\bibfnamefont {E.}~\bibnamefont
  {{Lichko}}}, \bibinfo {author} {\bibfnamefont {D.}~\bibnamefont
  {{Caprioli}}}, \bibinfo {author} {\bibfnamefont {B.}~\bibnamefont
  {{Schroer}}},\ and\ \bibinfo {author} {\bibfnamefont {S.}~\bibnamefont
  {{Gupta}}},\ }\bibfield  {title} {\bibinfo {title} {{Understanding Streaming
  Instabilities in the Limit of High Cosmic-Ray Current Density}},\ }\href
  {https://doi.org/10.3847/1538-4357/adadf5} {\bibfield  {journal} {\bibinfo
  {journal} {\apj}\ }\textbf {\bibinfo {volume} {980}},\ \bibinfo {eid} {240}
  (\bibinfo {year} {2025})},\ \Eprint {https://arxiv.org/abs/2411.05704}
  {arXiv:2411.05704 [astro-ph.HE]} \BibitemShut {NoStop}%
\bibitem [{\citenamefont {{Caprioli}}\ \emph {et~al.}(2020)\citenamefont
  {{Caprioli}}, \citenamefont {{Haggerty}},\ and\ \citenamefont
  {{Blasi}}}]{caprioli+20}%
  \BibitemOpen
  \bibfield  {author} {\bibinfo {author} {\bibfnamefont {D.}~\bibnamefont
  {{Caprioli}}}, \bibinfo {author} {\bibfnamefont {C.~C.}\ \bibnamefont
  {{Haggerty}}},\ and\ \bibinfo {author} {\bibfnamefont {P.}~\bibnamefont
  {{Blasi}}},\ }\bibfield  {title} {\bibinfo {title} {{Kinetic Simulations of
  Cosmic-Ray-modified Shocks. II. Particle Spectra}},\ }\href
  {https://doi.org/10.3847/1538-4357/abbe05} {\bibfield  {journal} {\bibinfo
  {journal} {\apj}\ }\textbf {\bibinfo {volume} {905}},\ \bibinfo {eid} {2}
  (\bibinfo {year} {2020})},\ \Eprint {https://arxiv.org/abs/2009.00007}
  {arXiv:2009.00007 [astro-ph.HE]} \BibitemShut {NoStop}%
\bibitem [{\citenamefont {{Drury}}(1983)}]{drury83}%
  \BibitemOpen
  \bibfield  {author} {\bibinfo {author} {\bibfnamefont {L.~O.}\ \bibnamefont
  {{Drury}}},\ }\bibfield  {title} {\bibinfo {title} {{REVIEW ARTICLE: An
  introduction to the theory of diffusive shock acceleration of energetic
  particles in tenuous plasmas}},\ }\href
  {https://doi.org/10.1088/0034-4885/46/8/002} {\bibfield  {journal} {\bibinfo
  {journal} {Reports on Progress in Physics}\ }\textbf {\bibinfo {volume}
  {46}},\ \bibinfo {pages} {973} (\bibinfo {year} {1983})}\BibitemShut
  {NoStop}%
\bibitem [{\citenamefont {Blasi}\ \emph {et~al.}(2007)\citenamefont {Blasi},
  \citenamefont {Amato},\ and\ \citenamefont {Caprioli}}]{blasi+07}%
  \BibitemOpen
  \bibfield  {author} {\bibinfo {author} {\bibfnamefont {P.}~\bibnamefont
  {Blasi}}, \bibinfo {author} {\bibfnamefont {E.}~\bibnamefont {Amato}},\ and\
  \bibinfo {author} {\bibfnamefont {D.}~\bibnamefont {Caprioli}},\ }\bibfield
  {title} {\bibinfo {title} {The maximum momentum of particles accelerated at
  cosmic ray modified shocks},\ }\href
  {https://doi.org/10.1111/j.1365-2966.2006.11412.x} {\bibfield  {journal}
  {\bibinfo  {journal} {\mnras}\ }\textbf {\bibinfo {volume} {375}},\ \bibinfo
  {pages} {1471} (\bibinfo {year} {2007})}\BibitemShut {NoStop}%
\bibitem [{\citenamefont {{Caprioli}}\ and\ \citenamefont
  {{Spitkovsky}}(2014{\natexlab{c}})}]{caprioli+14c}%
  \BibitemOpen
  \bibfield  {author} {\bibinfo {author} {\bibfnamefont {D.}~\bibnamefont
  {{Caprioli}}}\ and\ \bibinfo {author} {\bibfnamefont {A.}~\bibnamefont
  {{Spitkovsky}}},\ }\bibfield  {title} {\bibinfo {title} {{Simulations of Ion
  Acceleration at Non-relativistic Shocks. III. Particle Diffusion}},\ }\href
  {https://doi.org/10.1088/0004-637X/794/1/47} {\bibfield  {journal} {\bibinfo
  {journal} {\apj}\ }\textbf {\bibinfo {volume} {794}},\ \bibinfo {eid} {47}
  (\bibinfo {year} {2014}{\natexlab{c}})},\ \Eprint
  {https://arxiv.org/abs/1407.2261} {arXiv:1407.2261 [astro-ph.HE]}
  \BibitemShut {NoStop}%
\bibitem [{\citenamefont {{Schwartz}}\ \emph {et~al.}(1983)\citenamefont
  {{Schwartz}}, \citenamefont {{Thomsen}},\ and\ \citenamefont
  {{Gosling}}}]{schwartz+83}%
  \BibitemOpen
  \bibfield  {author} {\bibinfo {author} {\bibfnamefont {S.~J.}\ \bibnamefont
  {{Schwartz}}}, \bibinfo {author} {\bibfnamefont {M.~F.}\ \bibnamefont
  {{Thomsen}}},\ and\ \bibinfo {author} {\bibfnamefont {J.~T.}\ \bibnamefont
  {{Gosling}}},\ }\bibfield  {title} {\bibinfo {title} {{Ions upstream of the
  earth's bow shock - A theoretical comparison of alternative source
  populations}},\ }\href {https://doi.org/10.1029/JA088iA03p02039} {\bibfield
  {journal} {\bibinfo  {journal} {\jgr}\ }\textbf {\bibinfo {volume} {88}},\
  \bibinfo {pages} {2039} (\bibinfo {year} {1983})}\BibitemShut {NoStop}%
\bibitem [{\citenamefont {{Krauss-Varban}}\ and\ \citenamefont
  {{Wu}}(1989)}]{krauss-varban+89}%
  \BibitemOpen
  \bibfield  {author} {\bibinfo {author} {\bibfnamefont {D.}~\bibnamefont
  {{Krauss-Varban}}}\ and\ \bibinfo {author} {\bibfnamefont {C.~S.}\
  \bibnamefont {{Wu}}},\ }\bibfield  {title} {\bibinfo {title} {{Fast Fermi and
  gradient drift acceleration of electrons at nearly perpendicular
  collisionless shocks}},\ }\href {https://doi.org/10.1029/JA094iA11p15367}
  {\bibfield  {journal} {\bibinfo  {journal} {\jgr}\ }\textbf {\bibinfo
  {volume} {94}},\ \bibinfo {pages} {15367} (\bibinfo {year}
  {1989})}\BibitemShut {NoStop}%
\bibitem [{\citenamefont {{Mann}}\ \emph {et~al.}(2006)\citenamefont {{Mann}},
  \citenamefont {{Aurass}},\ and\ \citenamefont {{Warmuth}}}]{mann+06}%
  \BibitemOpen
  \bibfield  {author} {\bibinfo {author} {\bibfnamefont {G.}~\bibnamefont
  {{Mann}}}, \bibinfo {author} {\bibfnamefont {H.}~\bibnamefont {{Aurass}}},\
  and\ \bibinfo {author} {\bibfnamefont {A.}~\bibnamefont {{Warmuth}}},\
  }\bibfield  {title} {\bibinfo {title} {{Electron acceleration by the
  reconnection outflow shock during solar flares}},\ }\href
  {https://doi.org/10.1051/0004-6361:20064990} {\bibfield  {journal} {\bibinfo
  {journal} {\aap}\ }\textbf {\bibinfo {volume} {454}},\ \bibinfo {pages} {969}
  (\bibinfo {year} {2006})}\BibitemShut {NoStop}%
\bibitem [{\citenamefont {{Bohdan}}\ \emph {et~al.}(2022)\citenamefont
  {{Bohdan}}, \citenamefont {{Weidl}}, \citenamefont {{Morris}},\ and\
  \citenamefont {{Pohl}}}]{bohdan+22}%
  \BibitemOpen
  \bibfield  {author} {\bibinfo {author} {\bibfnamefont {A.}~\bibnamefont
  {{Bohdan}}}, \bibinfo {author} {\bibfnamefont {M.~S.}\ \bibnamefont
  {{Weidl}}}, \bibinfo {author} {\bibfnamefont {P.~J.}\ \bibnamefont
  {{Morris}}},\ and\ \bibinfo {author} {\bibfnamefont {M.}~\bibnamefont
  {{Pohl}}},\ }\bibfield  {title} {\bibinfo {title} {{The electron foreshock at
  high-Mach-number non-relativistic oblique shocks}},\ }\href
  {https://doi.org/10.1063/5.0084544} {\bibfield  {journal} {\bibinfo
  {journal} {Physics of Plasmas}\ }\textbf {\bibinfo {volume} {29}},\ \bibinfo
  {eid} {052301} (\bibinfo {year} {2022})},\ \Eprint
  {https://arxiv.org/abs/2204.05652} {arXiv:2204.05652 [physics.plasm-ph]}
  \BibitemShut {NoStop}%
\bibitem [{\citenamefont {{Raymond}}\ \emph {et~al.}(2023)\citenamefont
  {{Raymond}}, \citenamefont {{Ghavamian}}, \citenamefont {{Bohdan}},
  \citenamefont {{Ryu}}, \citenamefont {{Niemiec}}, \citenamefont {{Sironi}},
  \citenamefont {{Tran}}, \citenamefont {{Amato}}, \citenamefont {{Hoshino}},
  \citenamefont {{Pohl}}, \citenamefont {{Amano}},\ and\ \citenamefont
  {{Fiuza}}}]{raymond+23}%
  \BibitemOpen
  \bibfield  {author} {\bibinfo {author} {\bibfnamefont {J.~C.}\ \bibnamefont
  {{Raymond}}}, \bibinfo {author} {\bibfnamefont {P.}~\bibnamefont
  {{Ghavamian}}}, \bibinfo {author} {\bibfnamefont {A.}~\bibnamefont
  {{Bohdan}}}, \bibinfo {author} {\bibfnamefont {D.}~\bibnamefont {{Ryu}}},
  \bibinfo {author} {\bibfnamefont {J.}~\bibnamefont {{Niemiec}}}, \bibinfo
  {author} {\bibfnamefont {L.}~\bibnamefont {{Sironi}}}, \bibinfo {author}
  {\bibfnamefont {A.}~\bibnamefont {{Tran}}}, \bibinfo {author} {\bibfnamefont
  {E.}~\bibnamefont {{Amato}}}, \bibinfo {author} {\bibfnamefont
  {M.}~\bibnamefont {{Hoshino}}}, \bibinfo {author} {\bibfnamefont
  {M.}~\bibnamefont {{Pohl}}}, \bibinfo {author} {\bibfnamefont
  {T.}~\bibnamefont {{Amano}}},\ and\ \bibinfo {author} {\bibfnamefont
  {F.}~\bibnamefont {{Fiuza}}},\ }\bibfield  {title} {\bibinfo {title}
  {{Electron-Ion Temperature Ratio in Astrophysical Shocks}},\ }\href
  {https://doi.org/10.3847/1538-4357/acc528} {\bibfield  {journal} {\bibinfo
  {journal} {\apj}\ }\textbf {\bibinfo {volume} {949}},\ \bibinfo {eid} {50}
  (\bibinfo {year} {2023})},\ \Eprint {https://arxiv.org/abs/2303.08849}
  {arXiv:2303.08849 [astro-ph.GA]} \BibitemShut {NoStop}%
\bibitem [{\citenamefont {{Vanthieghem}}\ \emph {et~al.}(2024)\citenamefont
  {{Vanthieghem}}, \citenamefont {{Tsiolis}}, \citenamefont {{Spitkovsky}},
  \citenamefont {{Todo}}, \citenamefont {{Sekiguchi}},\ and\ \citenamefont
  {{Fiuza}}}]{vanthieghem+24}%
  \BibitemOpen
  \bibfield  {author} {\bibinfo {author} {\bibfnamefont {A.}~\bibnamefont
  {{Vanthieghem}}}, \bibinfo {author} {\bibfnamefont {V.}~\bibnamefont
  {{Tsiolis}}}, \bibinfo {author} {\bibfnamefont {A.}~\bibnamefont
  {{Spitkovsky}}}, \bibinfo {author} {\bibfnamefont {Y.}~\bibnamefont
  {{Todo}}}, \bibinfo {author} {\bibfnamefont {K.}~\bibnamefont
  {{Sekiguchi}}},\ and\ \bibinfo {author} {\bibfnamefont {F.}~\bibnamefont
  {{Fiuza}}},\ }\bibfield  {title} {\bibinfo {title} {{Electron Heating in High
  Mach Number Collisionless Shocks}},\ }\href
  {https://doi.org/10.1103/PhysRevLett.132.265201} {\bibfield  {journal}
  {\bibinfo  {journal} {\prl}\ }\textbf {\bibinfo {volume} {132}},\ \bibinfo
  {eid} {265201} (\bibinfo {year} {2024})},\ \Eprint
  {https://arxiv.org/abs/2405.09618} {arXiv:2405.09618 [astro-ph.HE]}
  \BibitemShut {NoStop}%
\bibitem [{\citenamefont {{Morlino}}\ \emph {et~al.}(2010)\citenamefont
  {{Morlino}}, \citenamefont {{Amato}}, \citenamefont {{Blasi}},\ and\
  \citenamefont {{Caprioli}}}]{morlino+10}%
  \BibitemOpen
  \bibfield  {author} {\bibinfo {author} {\bibfnamefont {G.}~\bibnamefont
  {{Morlino}}}, \bibinfo {author} {\bibfnamefont {E.}~\bibnamefont {{Amato}}},
  \bibinfo {author} {\bibfnamefont {P.}~\bibnamefont {{Blasi}}},\ and\ \bibinfo
  {author} {\bibfnamefont {D.}~\bibnamefont {{Caprioli}}},\ }\bibfield  {title}
  {\bibinfo {title} {{Spatial structure of X-ray filaments in SN 1006}},\
  }\href {https://doi.org/10.1111/j.1745-3933.2010.00851.x} {\bibfield
  {journal} {\bibinfo  {journal} {\mnras}\ }\textbf {\bibinfo {volume} {405}},\
  \bibinfo {pages} {L21} (\bibinfo {year} {2010})},\ \Eprint
  {https://arxiv.org/abs/0912.2972} {arXiv:0912.2972 [astro-ph.HE]}
  \BibitemShut {NoStop}%
\bibitem [{\citenamefont {{Merten}}\ \emph {et~al.}(2017)\citenamefont
  {{Merten}}, \citenamefont {{Becker Tjus}}, \citenamefont {{Eichmann}},\ and\
  \citenamefont {{Dettmar}}}]{merten+17}%
  \BibitemOpen
  \bibfield  {author} {\bibinfo {author} {\bibfnamefont {L.}~\bibnamefont
  {{Merten}}}, \bibinfo {author} {\bibfnamefont {J.}~\bibnamefont {{Becker
  Tjus}}}, \bibinfo {author} {\bibfnamefont {B.}~\bibnamefont {{Eichmann}}},\
  and\ \bibinfo {author} {\bibfnamefont {R.-J.}\ \bibnamefont {{Dettmar}}},\
  }\bibfield  {title} {\bibinfo {title} {{On the non-thermal electron-to-proton
  ratio at cosmic ray acceleration sites}},\ }\href
  {https://doi.org/10.1016/j.astropartphys.2017.02.007} {\bibfield  {journal}
  {\bibinfo  {journal} {Astroparticle Physics}\ }\textbf {\bibinfo {volume}
  {90}},\ \bibinfo {pages} {75} (\bibinfo {year} {2017})},\ \Eprint
  {https://arxiv.org/abs/1702.07523} {arXiv:1702.07523 [astro-ph.HE]}
  \BibitemShut {NoStop}%
\bibitem [{\citenamefont {Battiston}(2020)}]{battiston20}%
  \BibitemOpen
  \bibfield  {author} {\bibinfo {author} {\bibfnamefont {R.}~\bibnamefont
  {Battiston}},\ }\bibfield  {title} {\bibinfo {title} {High precision cosmic
  ray physics with ams-02 on the international space station},\ }\href
  {https://doi.org/10.1007/s40766-020-00007-2} {\bibfield  {journal} {\bibinfo
  {journal} {La Rivista del Nuovo Cimento}\ }\textbf {\bibinfo {volume} {43}},\
  \bibinfo {pages} {319} (\bibinfo {year} {2020})}\BibitemShut {NoStop}%
\bibitem [{\citenamefont {{Raptis}}\ \emph {et~al.}(2025)\citenamefont
  {{Raptis}}, \citenamefont {{Lalti}}, \citenamefont {{Lindberg}},
  \citenamefont {{Turner}}, \citenamefont {{Caprioli}},\ and\ \citenamefont
  {{Burch}}}]{raptis+25}%
  \BibitemOpen
  \bibfield  {author} {\bibinfo {author} {\bibfnamefont {S.}~\bibnamefont
  {{Raptis}}}, \bibinfo {author} {\bibfnamefont {A.}~\bibnamefont {{Lalti}}},
  \bibinfo {author} {\bibfnamefont {M.}~\bibnamefont {{Lindberg}}}, \bibinfo
  {author} {\bibfnamefont {D.~L.}\ \bibnamefont {{Turner}}}, \bibinfo {author}
  {\bibfnamefont {D.}~\bibnamefont {{Caprioli}}},\ and\ \bibinfo {author}
  {\bibfnamefont {J.~L.}\ \bibnamefont {{Burch}}},\ }\bibfield  {title}
  {\bibinfo {title} {{Revealing an unexpectedly low electron injection
  threshold via reinforced shock acceleration}},\ }\href
  {https://doi.org/10.1038/s41467-024-55641-9} {\bibfield  {journal} {\bibinfo
  {journal} {Nature Communications}\ }\textbf {\bibinfo {volume} {16}},\
  \bibinfo {eid} {488} (\bibinfo {year} {2025})},\ \Eprint
  {https://arxiv.org/abs/2502.10643} {arXiv:2502.10643 [astro-ph.HE]}
  \BibitemShut {NoStop}%
\bibitem [{\citenamefont {{Fulat}}\ \emph {et~al.}(2023)\citenamefont
  {{Fulat}}, \citenamefont {{Bohdan}}, \citenamefont {{Torralba Paz}},\ and\
  \citenamefont {{Pohl}}}]{karol+23}%
  \BibitemOpen
  \bibfield  {author} {\bibinfo {author} {\bibfnamefont {K.}~\bibnamefont
  {{Fulat}}}, \bibinfo {author} {\bibfnamefont {A.}~\bibnamefont {{Bohdan}}},
  \bibinfo {author} {\bibfnamefont {G.}~\bibnamefont {{Torralba Paz}}},\ and\
  \bibinfo {author} {\bibfnamefont {M.}~\bibnamefont {{Pohl}}},\ }\bibfield
  {title} {\bibinfo {title} {{Kinetic Simulations of Nonrelativistic
  High-mach-number Perpendicular Shocks Propagating in a Turbulent Medium}},\
  }\href {https://doi.org/10.3847/1538-4357/ad04dc} {\bibfield  {journal}
  {\bibinfo  {journal} {\apj}\ }\textbf {\bibinfo {volume} {959}},\ \bibinfo
  {eid} {119} (\bibinfo {year} {2023})},\ \Eprint
  {https://arxiv.org/abs/2310.14388} {arXiv:2310.14388 [physics.plasm-ph]}
  \BibitemShut {NoStop}%
\bibitem [{\citenamefont {{Kirk}}\ \emph {et~al.}(1996)\citenamefont {{Kirk}},
  \citenamefont {{Duffy}},\ and\ \citenamefont {{Gallant}}}]{kirk+96}%
  \BibitemOpen
  \bibfield  {author} {\bibinfo {author} {\bibfnamefont {J.~G.}\ \bibnamefont
  {{Kirk}}}, \bibinfo {author} {\bibfnamefont {P.}~\bibnamefont {{Duffy}}},\
  and\ \bibinfo {author} {\bibfnamefont {Y.~A.}\ \bibnamefont {{Gallant}}},\
  }\bibfield  {title} {\bibinfo {title} {{Stochastic particle acceleration at
  shocks in the presence of braided magnetic fields.}},\ }\href@noop {}
  {\bibfield  {journal} {\bibinfo  {journal} {\aap}\ }\textbf {\bibinfo
  {volume} {314}},\ \bibinfo {pages} {1010} (\bibinfo {year} {1996})},\ \Eprint
  {https://arxiv.org/abs/astro-ph/9604056} {astro-ph/9604056} \BibitemShut
  {NoStop}%
\bibitem [{\citenamefont {Bell}\ \emph {et~al.}(2011)\citenamefont {Bell},
  \citenamefont {Schure},\ and\ \citenamefont {Reville}}]{bell+11}%
  \BibitemOpen
  \bibfield  {author} {\bibinfo {author} {\bibfnamefont {A.~R.}\ \bibnamefont
  {Bell}}, \bibinfo {author} {\bibfnamefont {K.~M.}\ \bibnamefont {Schure}},\
  and\ \bibinfo {author} {\bibfnamefont {B.}~\bibnamefont {Reville}},\
  }\bibfield  {title} {\bibinfo {title} {Cosmic ray acceleration at oblique
  shocks},\ }\href {https://doi.org/10.1111/j.1365-2966.2011.19571.x}
  {\bibfield  {journal} {\bibinfo  {journal} {\mnras}\ }\textbf {\bibinfo
  {volume} {418}},\ \bibinfo {pages} {1208} (\bibinfo {year} {2011})},\ \Eprint
  {https://arxiv.org/abs/1108.0582} {arXiv:1108.0582 [astro-ph.HE]}
  \BibitemShut {NoStop}%
\bibitem [{\citenamefont {{Haggerty}}\ and\ \citenamefont
  {{Caprioli}}(2020)}]{haggerty+20}%
  \BibitemOpen
  \bibfield  {author} {\bibinfo {author} {\bibfnamefont {C.~C.}\ \bibnamefont
  {{Haggerty}}}\ and\ \bibinfo {author} {\bibfnamefont {D.}~\bibnamefont
  {{Caprioli}}},\ }\bibfield  {title} {\bibinfo {title} {{Kinetic Simulations
  of Cosmic-Ray-modified Shocks. I. Hydrodynamics}},\ }\href
  {https://doi.org/10.3847/1538-4357/abbe06} {\bibfield  {journal} {\bibinfo
  {journal} {\apj}\ }\textbf {\bibinfo {volume} {905}},\ \bibinfo {eid} {1}
  (\bibinfo {year} {2020})},\ \Eprint {https://arxiv.org/abs/2008.12308}
  {arXiv:2008.12308 [astro-ph.HE]} \BibitemShut {NoStop}%
\bibitem [{\citenamefont {{Longair}}(2011)}]{longair11}%
  \BibitemOpen
  \bibfield  {author} {\bibinfo {author} {\bibfnamefont {M.~S.}\ \bibnamefont
  {{Longair}}},\ }\href@noop {} {\emph {\bibinfo {title} {{High Energy
  Astrophysics}}}}\ (\bibinfo {year} {2011})\BibitemShut {NoStop}%
\bibitem [{\citenamefont {{Caprioli}}(2023)}]{caprioli23}%
  \BibitemOpen
  \bibfield  {author} {\bibinfo {author} {\bibfnamefont {D.}~\bibnamefont
  {{Caprioli}}},\ }\bibfield  {title} {\bibinfo {title} {{Particle Acceleration
  at Shocks: An Introduction}},\ }\href
  {https://doi.org/10.48550/arXiv.2307.00284} {\bibfield  {journal} {\bibinfo
  {journal} {arXiv e-prints}\ ,\ \bibinfo {eid} {arXiv:2307.00284}} (\bibinfo
  {year} {2023})},\ \Eprint {https://arxiv.org/abs/2307.00284}
  {arXiv:2307.00284 [astro-ph.HE]} \BibitemShut {NoStop}%
\bibitem [{\citenamefont {{de Hoffmann}}\ and\ \citenamefont
  {{Teller}}(1950)}]{HT50}%
  \BibitemOpen
  \bibfield  {author} {\bibinfo {author} {\bibfnamefont {F.}~\bibnamefont {{de
  Hoffmann}}}\ and\ \bibinfo {author} {\bibfnamefont {E.}~\bibnamefont
  {{Teller}}},\ }\bibfield  {title} {\bibinfo {title} {{Magneto-Hydrodynamic
  Shocks}},\ }\href {https://doi.org/10.1103/PhysRev.80.692} {\bibfield
  {journal} {\bibinfo  {journal} {Physical Review}\ }\textbf {\bibinfo {volume}
  {80}},\ \bibinfo {pages} {692} (\bibinfo {year} {1950})}\BibitemShut
  {NoStop}%
\end{thebibliography}%

\clearpage
\appendix
\section{Particle Acceleration at Shock}\label{app:A}
%
\subsection{Momentum Increment in DSA Cycle}
\begin{figure}[ht!]
\centering
\includegraphics[width=2.9in]{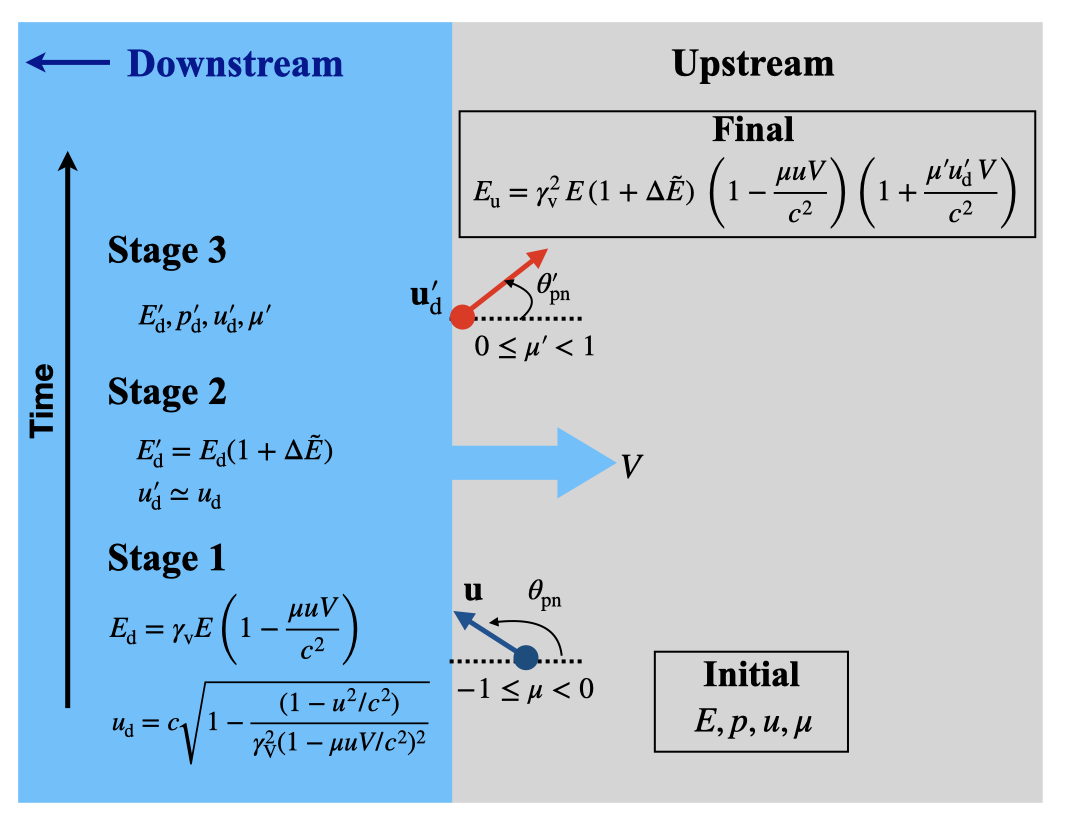}
\caption{
Schematic presentation of different stages of electron outrunning the shock observed from the upstream frame.
}\label{fig:sche}
\end{figure}

Consider an upstream particle with speed $u$ that approaches a nonrelativistic shock at an angle of incidence $\theta_{\rm pn}$ relative to shock normal, and then returns upstream with a different one, $\theta_{\rm pn}^{\rm \prime}$, as illustrated in Fig.~\ref{fig:sche}.
The classical derivation suggests that if $u\rightarrow c$, then
the average momentum gain in a complete cycle (upstream -- reflector -- upstream) is $\Delta p/p = 4 V/3c$ \citep[][]{longair11,caprioli23},
where $V$ is the speed of the reflector in the upstream rest frame. 
Here we consider arbitrary momenta, including nonrelativistic $u$.

Let $E,p,$ and $u$ be the initial energy, momentum, and speed of a particle measured in the upstream rest frame and $\mu\equiv\cos \theta_{\rm pn}$;
for particles approaching the shock $-1\leq \mu< 0 $.
Using the Lorentz transformations (see also Fig.~\ref{fig:sche}), 
the final particle energy upstream reads:
\begin{equation}
    E_{\rm u} = \gamma_{\rm v}^2\,E\, (1+ \Delta \tilde{E})\, \left(1 - \frac{\mu u V}{c^2}\right)\left(1 + \frac{\mu^{\rm \prime} u_{\rm d}^{\rm \prime}\, V}{c^2}\right).
\end{equation}
Here $\Delta \tilde{E}\equiv \Delta E/E^{\prime}_{\rm d}$ is the energy change due to any non-ideal mechanisms.
For simplicity, we assume $\Delta \tilde{E} \ll 1$, much smaller than the overall energy gain in the Fermi cycle.

Taking the probability of a particle impinging on the shock at an angle $\theta_{\rm pn}\equiv \cos^{-1}(\mu)$ as $P(\theta_{\rm pn}) d\theta_{\rm pn}=\frac{1}{2} \cos\theta_{\rm pn} \,\sin\theta_{\rm pn}\,d\theta_{\rm pn}=\mu d\mu/2$ (as expected for a hot distribution, $u\gtrsim V$), the average fractional momentum gain, $\mathcal{G} \equiv \langle \Delta p/p\rangle_{\mu,\mu^{\prime}} $, is obtained as 
\begin{equation}\label{eq:fullgaina}
    \mathcal{G}  = \frac{c^2}{u^2}\left\langle\frac{E_{\rm u}-E}{E}\right\rangle = \frac{c^2}{u^2} 
    \frac{ \int_{-1}^{0} d\mu \mu  \int_{0}^{1} d\mu^{\rm \prime} \mu^{\rm \prime} (\frac{E_{\rm u}}{E}-1)}{\int_{-1}^{0} d\mu \mu  \int_{0}^{1} d\mu^{\rm \prime} \mu^{\rm \prime}}
\end{equation}
Assuming $V\ll c$, we obtain Eq. (\ref{eq:frac_gainp}). 

\begin{figure}[ht!]
\centering
\includegraphics[width=2.9in]{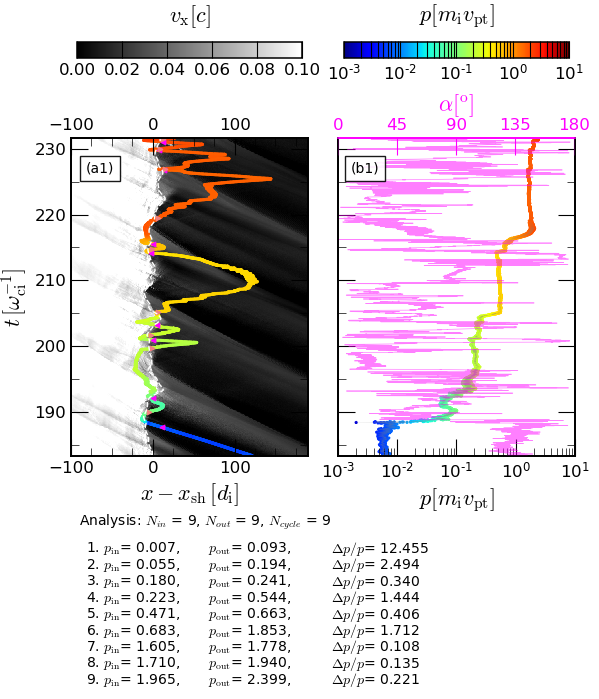}
\caption{
Demonstrating a method for calculating $\Delta p/p$ in PIC simulation. 
Descriptions are similar to Fig.~\ref{fig:M20trajectory}, with the background color representing the plasma x-velocity used to identify the instantaneous shock locations (the dashed white curve). 
The fractional momentum change, $\Delta p/p$, aligns with Eq. (\ref{eq:frac_gainp}).
}\label{fig:findingdelp}
\end{figure}

In Fig.~\ref{fig:findingdelp}, we illustrate a method for finding $\Delta p/p$ from the trajectory of an electron to produce Fig.~\ref{fig:M20gain}. 
To estimate the momentum of ingoing and outgoing particles, $p_{\rm in}$ and  $p_{\rm out}$, we first determine the shock location at different times. 
When the distance between the upstream particle and the shock at a given moment, $\Delta x_{\rm p}$, is smaller than $\Delta x_{\rm crit}$ (assuming $\Delta x_{\rm crit} = \Ma d_{\rm i}/4$), we classify the particle as an incoming particle and measure $p_{\rm in}$. 
If the particle subsequently returns to the upstream, such that $\Delta x_{\rm p}$ was initially $\ll \Delta x_{\rm crit}$ and then becomes larger with time, we label it as shock-reflected and measure $p_{\rm out}$. 
By repeating this method for each shock crossing, we estimate $\Delta p/p=p_{\rm out}/p_{\rm in}-1$.

\subsection{Escape Fraction}\label{appsec:escapefrac}
%
In our minimal model, the probability of particles remaining for acceleration is taken as $\mathcal{P}_{\rm rem} = 1 - \mathcal{F}_{\rm adv,x}/\mathcal{F}_{\rm in,x}$, where the flux of particles entering the shock from upstream ($\mathcal{F}_{\rm x,in}$) and then leaving the shock ($\mathcal{F}_{\rm x,out}$) are 
\begin{eqnarray}
     \mathcal{F}_{\rm in,x} & = & n\, c\,\int_{\rm -1}^{0} P(\mu)d\mu\,\sqrt{1-\frac{1- u^2/c^2}{\gamma_{\rm V}^2(1-\mu u V/c^2)^2}}\,\,,\ {\rm and} \nonumber \\
     \mathcal{F}_{\rm adv,x} & = &  n\, V_{\rm d}
\end{eqnarray}
respectively. Here $V_{\rm d}\equiv -(v_{\rm sh}-v_{\rm pt})=-v_{\rm pt}/(\mathcal{R}-1)$ is the downstream bulk speed relative to the shock. Assuming DSA particles are isotropic in local frame, i.e., $P(\mu) d\mu=\mu  d\mu/2$, we have $\mathcal{P}_{\rm rem} \simeq 1 - \frac{4}{\mathcal{R}-1}\left(v_{\rm pt}/u\right)$, as used in \S\ref{subsec:mimalmodel}. 
In the limit $u\rightarrow c$, $\mathcal{R}=4$, and $V=v_{\rm pt}\ll c$, $\mathcal{P}_{\rm rem}$ reduced to $\mathcal{P}_{\rm rem}=1- 4V/3c$, as predicted in the standard DSA theory.

By combining the momentum gain fraction, $\mathcal{G}\equiv \langle \Delta p/p\rangle$, with the probability of a particle remaining $\mathcal{P}_{\rm rem}$, the NT distribution emerges as a power law, $f(p)\propto p^{-q}$, with $q=3-\log(\mathcal{P}_{\rm rem})/\log(1+\langle \Delta p/p\rangle)$. 
Note that the superluminal magnetic structure upstream, or the drift of scattering centers relative to the plasma, can steepen the slope by reducing $\mathcal{P}_{\rm rem}$ and $\langle \Delta p/p\rangle$ \citep[e.g.,][]{caprioli+20,zekovic+24}.
\section{Electron Diffusion Across the Shock}\label{app:diffusion}
%
\subsection{Method}\label{app:methoddiff}
\begin{figure}[ht!]
\centering
\includegraphics[width=3.2in]{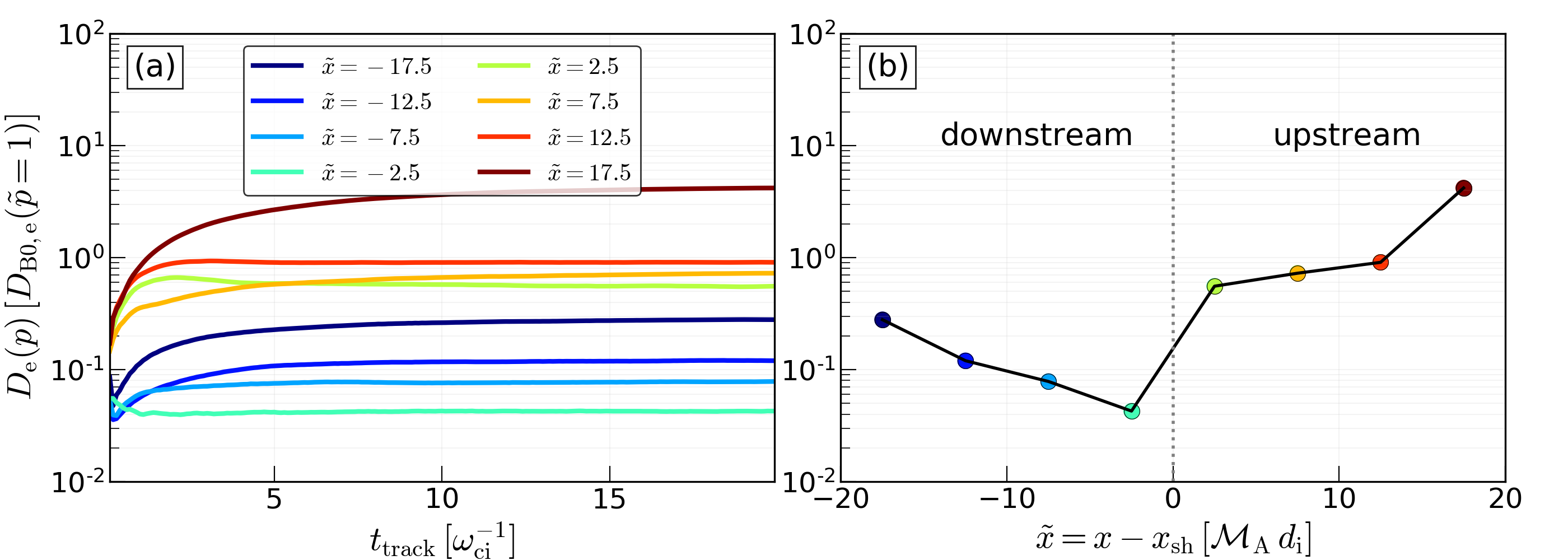}
\caption{
A method for finding $D_{\rm e}$ for $\Ma =20$ shock using a snapshot of the fields at $t=135\omega_{\rm ci}^{-1}$.
Panel (a) shows Eq. (\ref{eq:Dsim}) at eight different locations relative to the shock, while panel (b) presents the corresponding asymptotic values of $D_{\rm e}$.
}
\label{fig:tracdiff1}
\end{figure}

Here we outline a method, similar to \citep[][]{caprioli+14c}, for estimating $D_{\rm e}(p)$ across the shock to quantify $p_{\rm max,e}$ (\S\ref{subsec:maxp}).
We take a snapshot of the magnetic field from our PIC simulations and evolve test electrons using the Boris pusher, with a second-order shape function to estimate the electric and magnetic fields at the particle location and a fixed time step $\Delta t=0.045\,\omega_{\rm pe}^{-1}$.
To find $D(p)$ around a region $x_{\rm 0}$, we first select a domain $x_{\rm 0}\pm L$, where the length $2L$ is comparable to the precursor length $\sim D_{\rm B}(p)/v_{\rm sh}\equiv \tilde{p}^2 \mr /\gamma\, \Ma d_{\rm i}$, which is derived from Eq. (\ref{eq:Bohmdiff}).
We mirror the domain, making the total size $4L$, and apply periodic boundary condition to all directions to ensure particles experience continuous fields. Each cell, sized at $d_{\rm e}/10$ as in the PIC simulation, contains test particles.
We track particle displacement over time and derive the diffusion coefficient using Eq. (\ref{eq:Dsim}) \citep[see, e.g.,][]{caprioli+14c}, as shown in Fig.~\ref{fig:tracdiff1}.
\begin{eqnarray}\label{eq:Dsim}
    D_{\rm e}(p) =  \lim_{t\rightarrow \infty} \frac{1}{N\,t_{\rm track}}\sum_{\rm n=0}^{N-1} |x_{\rm n}(t_{\rm track}) - x_{\rm n}(0)|^2 .
\end{eqnarray} 
Fig.~\ref{fig:tracdiff1}(a) shows the time evolution of Eq. (\ref{eq:Dsim}).
For most curves, $D_{\rm e}$ reach an asymptotic value (i.e., diffusive regime); while a few of them do not, representing super-diffusion/streaming regime at far upstream.
Fig.~\ref{fig:tracdiff1}(b) represents the profiles of $D_{\rm e}$: in the region immediately ahead of shock $D_{\rm 1,e}\approx 0.6 D_{\rm B0}$ and behind the shock $D_{\rm 2,e}\approx D_{\rm 1,e}/10$.
Such a reduction in the electron diffusivity is due to large $\delta B/B$ fluctuations produced by the proton-driven streaming instabilities, as generally found in a high $\Ma $ shock \citep[see Fig.~2 in][]{gupta+24b}.
$\delta B/B$ decreases with increasing distance from the shock, causing an enhanced diffusivity in the far upstream.

\subsection{Evolution of Maximum Momentum}\label{app:diffusionmR}
%
\begin{figure}
\centering
\includegraphics[width=0.45\textwidth, trim=5px 0px 10px 0px, clip=true]{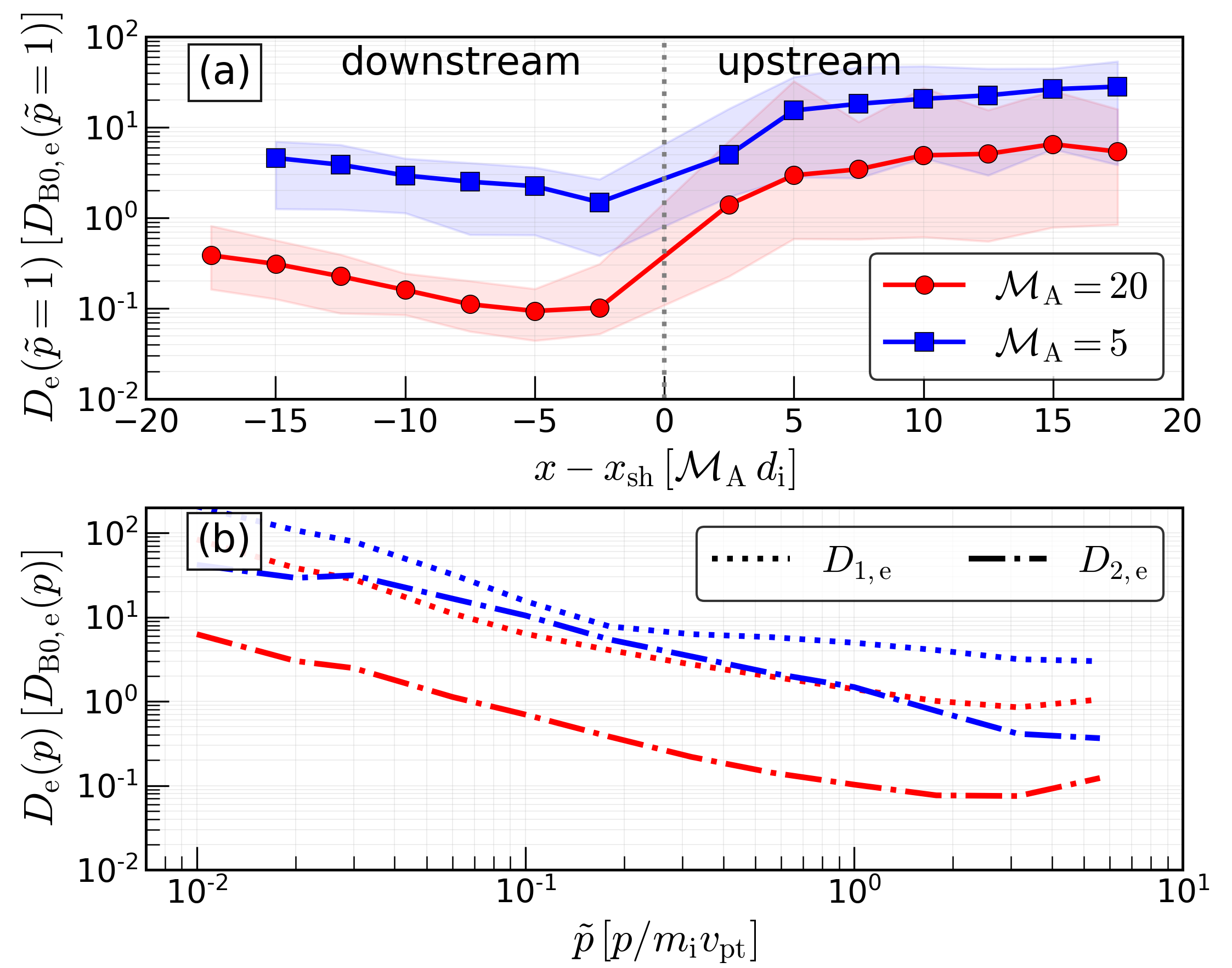}
\caption{
The electron diffusion coefficient, $D_{\rm e}(p)$, normalized to the Bohm diffusion coefficient of the initial magnetic field [Eq. (\ref{eq:Bohmdiff})]. 
The red and blue curves stand for $\Ma =20$ and $\Ma =5$ shock, respectively where $v_{\rm pt}/c=0.1$, $\Ms  =40$, and $\mr =100$.
Panel (a) shows the time-averaged spatial profile of $D_{\rm e}$ using $p=m_{\rm i}v_{\rm pt}$, with the shaded areas indicating temporal variations over $100-275\,\omega_{\rm ci}^{-1}$.
Panel (b) shows the time-averaged $D_{\rm e}$ as a function of $p$, with $D_{\rm e,1,2}$ corresponding to a region around $x=x_{\rm sh}\pm 2.5\,\Ma d_{\rm i}$, using the same color scheme as in panel (a).
}
\label{fig:tracdiff2}
\end{figure}

To quantify the momentum evolution of NT electrons in the $\Ma=20$ and $\Ma=5$ shocks, as discussed in \S \ref{subsec:maxp}, we compute the electron diffusion coefficient across the shock. This is done using the method described in Appendix~\ref{app:methoddiff}. Since the NT spectra at a given time depend on the history of turbulence across the shock, we use a time-averaged $D_{\rm e}(p)$ to find $p_{\rm max,e}$. 
The resulting diffusion coefficients are shown in Fig.~\ref{fig:tracdiff2}.

Fig.~\ref{fig:tracdiff2}(a) displays the spatial profile of $D_{\rm e}(p)$ at a fixed momentum $p=m_{\rm i}v_{\rm pt}$ (red for $\Ma =20$ and blue for $\Ma =5$).
Fig.~\ref{fig:tracdiff2}(b) represents the time-averaged $D_{\rm e}(p)$ at $x=x_{\rm sh}\pm 2.5\Ma d_{\rm i}$ as a function of $p$, showing the downstream $D_{\rm 2}$ is smaller than upstream, $D_{\rm 1}$.
For $\Ma =20$ shock (red curves in Fig.~\ref{fig:tracdiff2}), substituting $D_{\rm 1,e}\approx D_{\rm B0,e}$ and $D_{\rm 2,e}=0.1 D_{\rm B0,e}$ into Eq. (\ref{eq:pmax}) gives $p_{\rm max,e}/m_{\rm i}v_{\rm pt}\approx 9$, in good agreement with the maximum momentum obtained in kinetic simulations (see also Fig.~\ref{fig:M20trajectory}(b)).
For $\Ma =5$ (blue curves), a similar estimate returns $p_{\rm max,e}/m_{\rm i}v_{\rm pt}\approx 0.9$, as mentioned in \S \ref{subsec:maxp}.

\begin{figure}[ht!]
\centering
\includegraphics[width=3.4in]{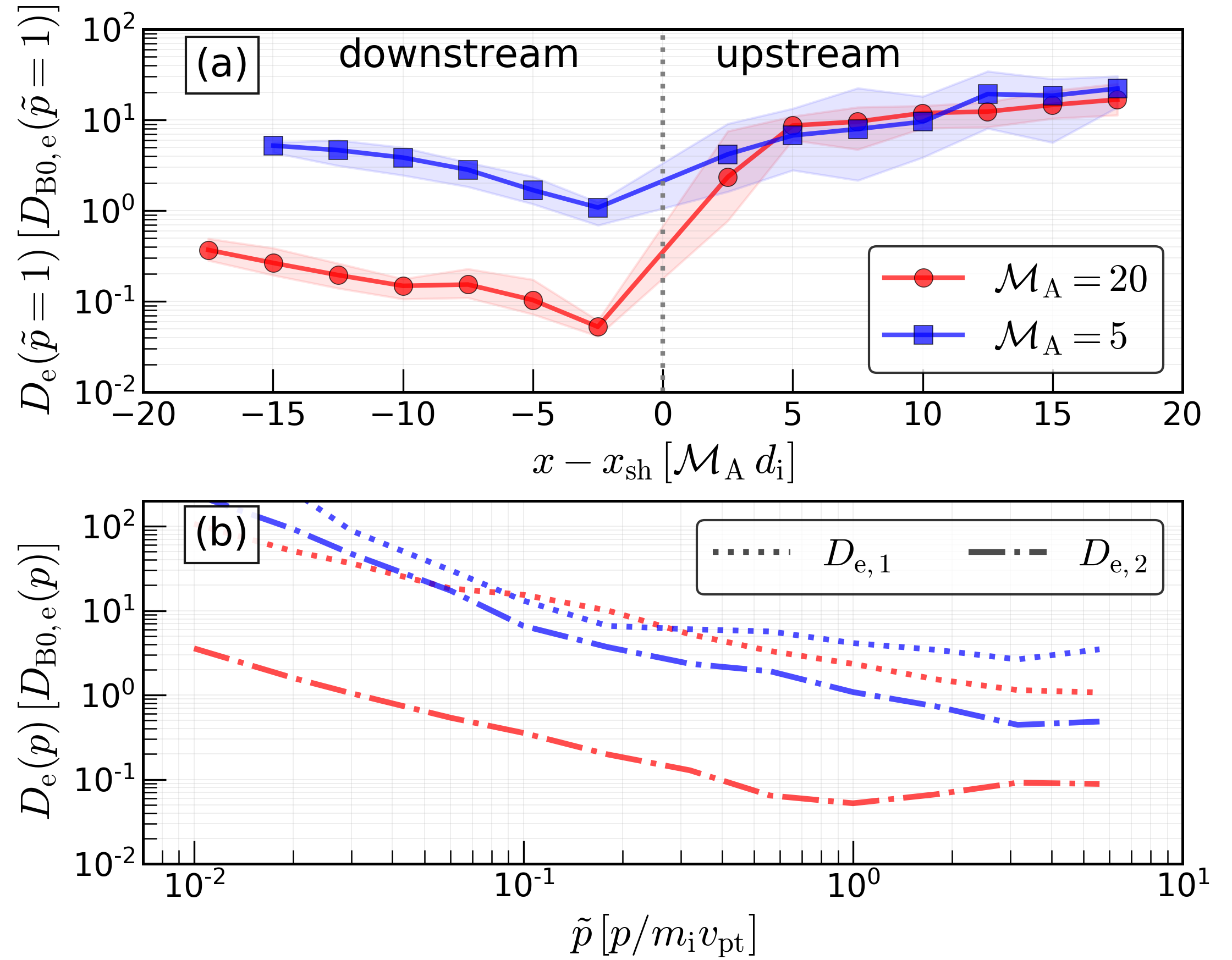}
\caption{
Same as Fig.~\ref{fig:tracdiff2}, but using $\mr =1836$.}
\label{fig:tracdiffmR}
\end{figure}
To verify that the conclusions from a reduced mass-ratio ($\mr =100$, \S\ref{subsec:maxp}) hold for a realistic mass-ratio ($\mr =1836$), we analyze electron diffusion in $\Ma =5$ and $\Ma =20$ shocks, but now they are run with $\mr =1836$, keeping all other parameters unchanged \citep[also see][]{gupta+24b}.
Fig.~\ref{fig:tracdiffmR}(a) shows the spatial profile of $D_{\rm e}$ for these tests, but averaged over $100-150\omega_{\rm ci}^{-1}$, as extending simulations with a realistic mass-ratio is computationally demanding.

We find that $D_{\rm 1,e}$ for the $\Ma  = 20$ \& $\mr =1836$ case (red curves) is slightly larger than that found in the $\mr  = 100$ case.
Such a difference arises because the linear stage (hence the saturation stage) of the proton-driven streaming instability begins later for $\mr  = 1836$.
This causes a smaller $\delta B/B$ \citep[see Appendix A in][]{gupta+24b}, leading to a larger $D_{\rm e}$ [Eq. (\ref{eq:Bohmdiff})] and a slightly smaller $p_{\rm max,e}$. 
For the $\Ma  = 5$ shock (blue curve), we find $p_{\rm max,e} \approx 0.8 m_{\rm i} v_{\rm pt}$, consistent with the results from the $\mr  = 100$ case, as $\delta B/B$ does not increase in either case due to a low $\Ma$. 
Both results agree with the spectra found in fully kinetic shock simulations \citep[see, e.g., Fig.~6 in][]{gupta+24b}.

\section{Reflection at an Idealized Shock}\label{app:syntheticshock}
%
\begin{figure}[ht!]
\centering
\includegraphics[width=3.4in]{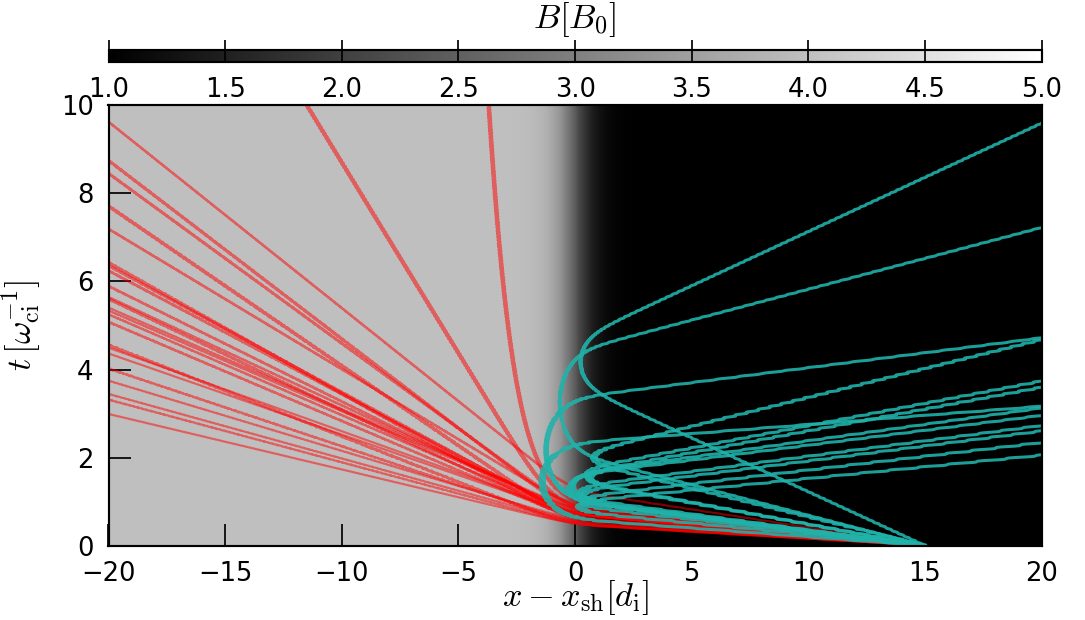}
\caption{
Trajectories of a few transmitted (red) and reflected (cyan) electrons in a prescribed shock test, where we have taken $v_{\rm pt} = 0.1\,c$, $\thetabn = 45^{\circ}$, $\mathcal{M}_{\rm s,e} = 2$, $L_{\rm sh} = 1d_{\rm i}$, $\mathcal{R}_{\rm B} = 4$, and proton-to-electron mass ratio $m_{\rm R}=100$.
}\label{fig:test_reflection}
\end{figure}

\begin{figure}[ht!]
\centering
\includegraphics[width=3.4in]{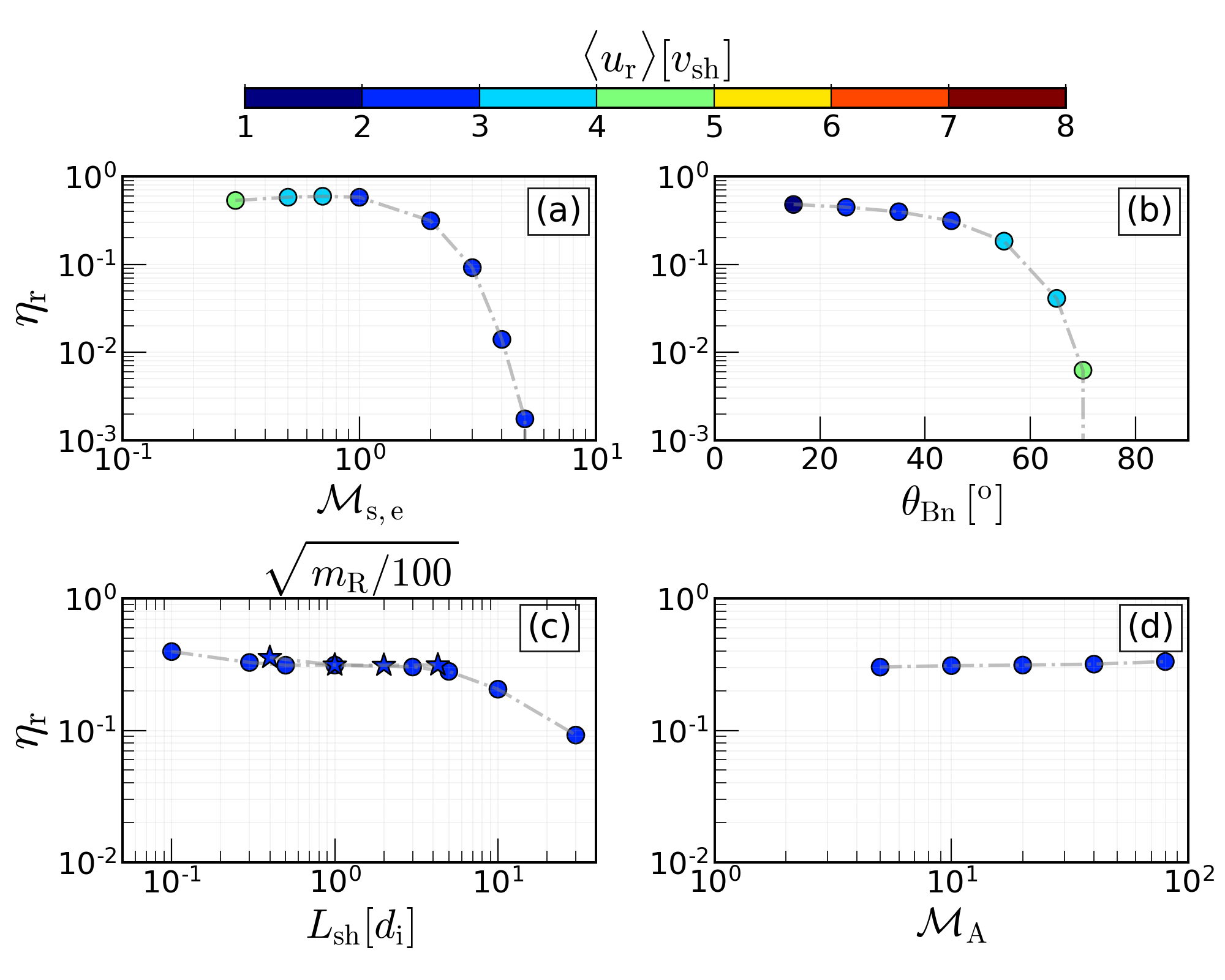}
\caption{
Quantification of the electron reflection fraction ($\eta_{\rm r}$) using prescribed shock simulations. Panels (a) and (b) show that $\eta_{\rm r}$ drops sharply for $\mathcal{M}_{\rm s,e} >3$ and at higher inclinations. Panel (c) indicates that varying the shock thickness--either by adjusting $L_{\rm sh}$ (bottom axis, circles) or the mass ratio (top axis, stars)--has little effect on $\eta_{\rm r}$. Panel (d) suggests a mild dependence on $\mathcal{M}_{\rm A}$. Notably, in all cases, the mean velocity of reflected electrons in the upstream frame is $1$–$4\,v_{\rm sh}$ (see top color bar), consistent with our full PIC simulations.
}
\label{fig:synref_cr}
\end{figure}
Here, we elaborate on the electron reflection discussed in \S\ref{subsec:mimalmodel}.
Electrons in the upstream observe the shock as a gradually increasing electromagnetic structure moving at a speed $v_{\rm sh}$. Since the Larmor radius of thermal electrons is typically smaller than the shock thickness, electrons can get reflected via a process called magnetic mirroring. However, unlike the classical case, here the mirror moves, and thus it contains non-zero motional electric fields.
This can be handled by choosing the de Hoffmann--Teller frame, which moves along the magnetic field with velocity ${\bf v}_{\rm HT}=v_{\rm sh}\sec\thetabn {\bf b}$ in the upstream rest frame \citep[][]{HT50,guo+14a}.
An electron can be reflected by the shock if its velocity components parallel and perpendicular to the shock normal satisfy the conditions:
\begin{eqnarray} \label{eq:condi-1}
   u_{\rm \parallel,i}^{HT}& = & \frac{u_{\rm \parallel,i} - v_{\rm HT}}{1- u_{\rm \parallel,i} v_{\rm HT}/c^2}  < 0 ,\ {\rm and} \\
\label{eq:condi-2}
  u_{\rm \perp,i}^{HT}& = & \frac{u_{\rm \perp,i} \sqrt{ 1 - (v_{\rm HT}/c)^2}}{1- u_{\rm \parallel,i} v_{\rm HT}/c^2} > \frac{u_{\rm \parallel,i}^{HT}}{\sqrt{\mathcal{R}_{\rm B}-1}} \, .
\end{eqnarray}
Here $\mathcal{R}_{\rm B}=B_{\rm 2}/B_{\rm 1}$ is the magnetic compression ratio across the shock.
Note that we use superscript to denote the reference frame of measurement, subscript to specify the quantity or associated frame, similar to \citep[][]{guo+14a}.  The quantities without superscripts are defined in the upstream frame.
Eqs. (\ref{eq:condi-1}) and (\ref{eq:condi-2}) can be used to obtained a lower bound on $u=\sqrt{ u_{\rm \parallel,i }^2+u_{\rm \perp,i }^2}$ with respect to $u_{\rm \parallel}$, which gives Eq. (\ref{eq:condi-3}).
This condition requires the effective sonic Mach number for electrons to be $\lesssim 3$.

We design a synthetic shock setup to validate the electron reflection condition.
We prescribe a shock using a step function  $\tanh(-x/L_{\rm sh})$, where $x$ is a distance relative to the shock and $L_{\rm sh}$ is the width of the shock transition, for the electro-magnetic profiles.
The smooth transition from a high state (downstream) to a low state (upstream) represents the shock surface.
We push test-particle thermal electrons and measure how many of them are reflected.   
Fig.~\ref{fig:test_reflection} shows typical transmitted and reflected electrons trajectories, while Fig.~\ref{fig:synref_cr} presents the reflection fraction ($\eta_{\rm refl}$). Fiducial parameters are $v_{\rm pt} = 0.1$, $\mathcal{R}_{\rm B} = 4$, $\Mse = 2$, $\thetabn = 45^\circ$, $L_{\rm sh} = d_{\rm i}$, $\Ma = 20$, and $m_{\rm R}=100$. Figure shows that quasi-parallel shocks reflect $\sim 10\%$ of electrons for $\Mse\lesssim 3$.

While the setup described above demonstrates magnetic mirroring of electrons, we show below that reflection alone may not be sufficient for an electron to outrun the shock and participate in DSA when the upstream turbulence (and $\thetabn$) evolves rapidly. Since electrons take finite time to get reflected (as shown in Fig~\ref{fig:test_reflection}), the shock can reform, and the local magnetic field inclination may change before electrons escape from the shock. To account for such effects, let $u_{\rm \parallel,r}$ and $u_{\rm \perp,r}$ be the velocity components of an electron in the upstream frame just after undergoing reflection in the HT frame; these are found as (see \citep[][]{guo+14a})
\begin{eqnarray}
    u_{\rm \parallel,r} &=& \frac{2 v_{\rm HT} - u_{\rm \parallel,i}(1+v_{\rm HT}^2/c^2)}{1- 2u_{\rm \parallel,i}v_{\rm HT}/c^2+v_{\rm HT}^2/c^2}\nonumber,\\
    u_{\rm \perp,r} &=& \frac{u_{\rm \perp,i}\sqrt{1-v_{\rm HT}^2/c^2}}{1- 2u_{\rm \parallel,i}v_{\rm HT}/c^2+v_{\rm HT}^2/c^2}.
\end{eqnarray}
 To find the corresponding shock normal velocity in the upstream frame, we define an orthogonal coordinate system with unit vectors ${\bf b}$, ${\bm \zeta}$ and ${\bm \xi}$ such that ${\bf b}$ is along the magnetic field. Then, the perpendicular velocity becomes ${\bf u}_{\rm \perp,r}=u_{\rm \perp,r}\cos(\tau_{\rm c}+\phi_{\rm r}) {\bm \zeta} + u_{\rm \perp,r}\sin(\tau_{\rm c}+\phi_{\rm r}) {\bm \xi}$, where $\phi_{\rm r}\in [0,2 \pi]$ and $\tau_{\rm c}=\omega_{\rm c}t$ is time normalized to the electron gyrofrequency.
Assuming the magnetic field lies in the $x-y$ plane, the shock normal (${\bf x}\equiv \cos\thetabn^{\rm \prime}\,{\bf b} - \sin \thetabn^{\rm \prime}\,{\bm \zeta}$) velocity of reflected electrons in the upstream frame is obtained as
\begin{equation}
    u_{\rm r,x} = u_{\rm \parallel,r} \cos\thetabn^{\rm \prime} - u_{\rm \perp,r}\cos(\tau_{\rm c}+\phi_{\rm r}) \sin\thetabn^{\rm \prime}, 
\end{equation}
where $\thetabn^{\rm \prime}\sim \cos^{-1}(B_{\rm x}/|B|)$ is the local inclination of the shock as seen by an electron immediately after the reflection. Note that $\thetabn^{\prime}$ differs from $\thetabn$ in the presence of strong upstream turbulence. Thus, the averaged velocity along the shock normal over a gyroperiod is
\begin{equation}
   \langle u_{\rm r,x} \rangle_{\rm \tau_{\rm c}}= u_{\rm \parallel,r} \cos\thetabn^{\rm \prime} - u_{\rm \perp,r}\sin\thetabn^{\rm \prime}\left[\sin(\tau_{\rm c}+\phi_{\rm r})-\sin\phi_{\rm r}\right].
\end{equation}
Reflected electrons can escape upstream and eventually be injected into DSA if $\langle u_{\rm r,x}\rangle_{\rm \tau_{\rm c}}$ exceeds the shock speed. This condition results in the injected fraction $\lesssim 1/2$ of reflected electrons. 

\end{document}